\documentclass[useAMS,usenatbib]{mn2e}
\usepackage{graphicx,times}

\title[Loss of halo triaxiality due to bar formation]{Loss of halo triaxiality
due to bar formation}

\author[R. E. G. Machado and E. Athanassoula]{Rubens E. G.
Machado$^{1,2}$\thanks{E-mail:
rgmachado@astro.iag.usp.br} and E. Athanassoula$^{1}$\\
$^{1}$Laboratoire d'Astrophysique de Marseille (LAM), UMR6110, 
CNRS/Universit\'e de Provence,\\ Technop\^ole de Marseille Etoile, 38 rue
Fr\'ed\'eric Joliot Curie,
13388 Marseille C\'edex 13, France\\
$^{2}$Instituto de Astronomia, Geof\'isica e Ci\^encias Atmosf\'ericas,
Universidade de S\~ao Paulo,\\
R. do Mat\~ao 1226, 05508-090 S\~ao Paulo, Brazil}

\begin{document}

\date{Accepted 2010 April 21. Received 2010 March 31; in original form 2010 February 21}

\pagerange{\pageref{firstpage}--\pageref{lastpage}} \pubyear{2010}

\maketitle

\label{firstpage}

\begin{abstract}

Cosmological $N$-body simulations indicate that the dark matter haloes of
galaxies should be generally triaxial. Yet, the presence of a baryonic disc is
believed to alter the shape of the haloes.
Here we aim to study how bar formation is affected by halo triaxiality and how,
in turn, the presence of the bar influences the shape of the halo.
We perform a set of collisionless $N$-body simulations of disc galaxies with
triaxial dark matter haloes, using elliptical discs as initial conditions. Such
discs are much closer to equilibrium with their haloes than circular ones, and
the ellipticity of the initial disc depends on the ellipticity of the halo
gravitational potential. For comparison, we also consider models with initially
circular discs, and find that the differences are very important. In all cases,
the mass of the disc is grown quasi-adiabatically within the haloes, but
the time-scale of growth is not very important. We study models of different
halo triaxialities and, to investigate the behaviour of the halo shape in the
absence of bar formation, we run models with different disc masses, halo
concentrations, disc velocity dispersions and also models where the disc shape
is kept artificially axisymmetric.
We find that the introduction of a massive disc, even if this is not circular,
causes the halo triaxiality to be partially diluted. Once the disc is fully
grown, a strong stellar bar develops within the halo that is still
non-axisymmetric, causing it to lose its remaining non-axisymmetry. In triaxial
haloes in which the parameters of the initial conditions are such that a bar
does not form, the halo is able to remain triaxial and the circularisation
of its shape on the plane of the disc is limited to the period of disc growth.
We conclude that part of the circularisation of the halo is due to disc growth,
but part must be attributed to the formation of a bar. Bars in the halo
component, which have been already found in axisymmetric haloes, are also found
in triaxial ones. We find that initially circular discs respond excessively to
the triaxial potential and become highly elongated. They also lose more angular
momentum than the initially elliptical discs and thus form stronger bars.
Because of that, the circularisation that their bars induce on their haloes is
also more rapid. We also analyse halo vertical shapes and observe that their
vertical flattenings remain considerable, meaning that the haloes become
approximately oblate by the end of the simulations. Finally, we also analyse the
kinematics of a subset of halo particles that rotate in disc-like manner. These
particles occupy a layer around the plane of the disc and their rotation is more
important in the spherical halo than in triaxial ones. We also find that, even
though the final shape of the halo is roughly independent of the initial shape,
the initially triaxial ones are able to retain the anisotropy of their velocity
dispersions.

\end{abstract}



\section{Introduction}
\label{sec:intro}

The vast majority of papers studying bar formation and evolution use
idealised galaxy models with an exponential, or near-exponential, disc and a
spherical halo. Yet, already in the nineties,
cosmological $N$-body simulations had shown that dark matter haloes are
generally non-spherical, with a tendency to be more prolate than
oblate \citep{Frenk1988, Dubinski1991, Warren1992, Cole1996}. Later
simulations \citep{Jing2002, BailinSteinmetz2005, Allgood2006, Novak+2006}
have a sufficiently large number of particles to allow an adequate
statistical analysis of the halo properties. Typically it is found
that such haloes have isodensity axis ratios of $b/a \sim 0.8$ and
$c/a \sim 0.6$, depending on the mass of the halo and on the details
of the cosmological simulations. In particular, more massive haloes
tend to be more triaxial (or rather, prolate). This is presumably due
to the fact that massive haloes undergo a larger number of merging
events, which take place non-isotropically, along preferred
directions linked to the filaments of the global large-scale
structure. Furthermore, the axis ratios show some radial dependence
and the triaxiality is usually found to increase towards the centre, as
seen by \citet*{Hayashi2007} who measured the shapes of the isopotential
surfaces of cosmological haloes.

Simulations of large scale structure used to be restricted to dark matter,
ignoring the baryonic component, due to the computational cost. More recently,
cosmological simulations that include gas and the treatment of several physical
processes (such as star formation, gas cooling, chemical enrichment and
supernova feedback) have been able to form discs. And it is found that haloes
tend to become axisymmetric due to the disc \citep{Kazantzidis2004,
Tissera+2009}. Thus a halo which contains a baryonic disc is expected to be less
triaxial than a pure dark matter halo. Indeed, observational constraints on halo
shape find that present-day haloes are very mildly elongated on the plane of the
disc, or even consistent with an axisymmetric potential
\citep{Trachternach2008}. This is also reflected on the statistics of disc
galaxy shapes (\citealt*{Lambas1992}; \citealt{Fasano1993};
\citealt{RixZaritsky1995}; \citealt{Ryden2004, Ryden2006}), whose ellipticities
are small compared to the ellipticities of haloes from cosmological
simulations. 

The shapes of discs are deformed by the aspheric halo potentials, such that in
the equilibrium configuration the discs should be generally elliptical. At the
same time, the presence of the disc acts to oppose the halo ellipticity on the
plane \citep{Jog2000, Bailin2007}, since the disc elongation is perpendicular
to the halo elongation. The interplay of the baryonic disc with its halo should
somehow reconcile the highly triaxial shapes of pure dark matter haloes from
cosmological simulations with the near circularity of present-day observed
galaxies.

Such disc galaxies may develop bars, whose dynamics must be influenced both by
the non-circular disc and the non-spherical halo. In order to investigate 
the effect of a triaxial halo on bar formation, \citet{GadottideSouza2003} 
performed $N$-body simulations of a spheroid embedded in a rigid triaxial 
halo potential. In their simulations, the spheroid was distorted into a 
bar-like structure, due to the halo triaxiality.

The effects of a cosmological setting were studied by \citet*{Curir+2006}, who
embedded circular discs in the haloes of cosmological simulations. They argue
that the large scale anisotropies of the mass distribution influence the bar
strengths and details of bar evolution, due to the continuous matter infall and
substructures in the halo.

With simulations of considerably higher mass resolution,
\citet*{BerentzenShlosmanJogee2006} examined the effect of mildly triaxial
haloes on bar evolution. In two of their three simulations with a live triaxial
halo, they found that the bar quickly dissolved. The third case had a very
massive and very mildly triaxial halo. In particular, its isopotential
axis ratio in the equatorial plane is about 0.9. It this last case the bar does
not dissolve.

\citet{BerentzenShlosman2006} performed $N$-body simulations in which they grew
circular seed discs in assembling triaxial haloes, within a
quasi-cosmological setting. The final shape of the halo depends on the mass of
the disc, but not on the timescale of its growth. They show that massive discs
completely wash out the halo prolateness and then develop long-lived bars,
whereas discs that contribute less to the rotation curve are less efficient in
axisymmetrising their haloes. They claim that in less massive discs, the bar
instability is damped by the halo triaxiality.

\citet*{HellerShlosmanAthanassoula2007a} investigated the formation of discs by
following the collapse of an isolated cosmological density perturbation. They
include star formation and stellar feedback in their simulations, so that a
baryonic disc forms inside the assembling dark matter halo. The halo triaxiality
is decreased in their models during disc growth. Bars that are formed early
decay within a few Gyr, but such bars are driven by the prolateness of the halo
and do not follow the usual bar instability evolution. Also, they find that the
tumbling of the triaxial halo figure is insignificant.

\citet{Widrow2008} used the adiabatic squeezing method of
\citet{HolleyBockelmann2001} to produce triaxial halo models, which he applied
to a study of F568-3. This work was focused on a study of the rotation curve of
this galaxy and did not discuss the reasons for the changes of the halo and disc
shapes. \citet{Debattista2008} carried out control simulations in which a rigid
disc was adiabatically first grown and then evaporated and found that the haloes
were substantially rounder when the disc was near full mass, but that they
returned to their initial shape after the disc was evaporated. The main goal of
this paper and of its sequel \citep{Valluri2009arXiv} was to search for the
changes of orbital structure that could account for the changes of halo shape
due to the baryonic component.

In the present work we continue along the lines of the above mentioned work and
particularly investigate bar formation and evolution in triaxial haloes and the
corresponding effects on the halo properties. Does the triaxiality of the halo
inhibit bar formation, or change drastically the bar properties? Or,
alternatively, are strong bars able to form inside triaxial haloes and then
cause them to lose their remaining triaxiality? We investigate different models
and different types of initial conditions. As initial conditions, we use
elliptical discs which are designed to be in equilibrium with the elliptical
potential of the halo and compare the results with those obtained with the more
straightforward, but out of equilibrium, initially circular discs. We also focus
on evaluating the separate effects of two factors that contribute to changing
the halo shape: the growth of the disc mass, and the formation of a bar. 

In Sect.~\ref{sec:initialconditions} we present our initial conditions and in
Sect.~\ref{sec:ECdisks} we discuss bar growth and compare the results in
initially axisymmetric and in initially elongated discs. In
Sects.~\ref{sec:coresizes} and \ref{sec:timescales} we consider different halo
core sizes and different time-scales for disc growth, respectively. In
Sect.~\ref{sec:nobars} we present a series of simulations, made in order to
distinguish how much of the evolution of the halo shape is due to the
introduction of the disc and how much to the growth of the bar. Vertical
shapes are the subject of Sect.~\ref{sec:covera}. In 
Sect.~\ref{sec:disklike} we consider the effect of triaxiality on halo
kinematics. Finally we summarise and conclude in Sect.~\ref{sec:conclusions}.

\begin{table}
\caption{Initial shapes of the haloes as given by the intermediate-to-major
($b/a$) and minor-to-major ($c/a$) axis ratios.}
\label{tb:halos}
\begin{center}
\begin{tabular}{l c c}
\hline
halo & $b/a$ & $c/a$\\
\hline
1  & 1.0 & 1.0 \\ 
2  & 0.8 & 0.6 \\
3-- & 0.7 & 0.5 \\ 
3  & 0.6 & 0.4 \\
3+ & 0.5 & 0.3 \\
\hline
\end{tabular}
\end{center}
\end{table}


\section{Initial conditions}
\label{sec:initialconditions}

\subsection{Halo initial conditions} \label{subsec:haloic}

\begin{table*}
\caption{Parameters of the initial conditions for all simulations: (1) model
name; (2) halo (see halo shapes in Table \ref{tb:halos}); (3) disc shape; (4)
disc mass; (5) halo mass; (6) halo core size; (7) Toomre parameter; (8) relative
orientation of the disc and halo major axes; (9) time-scale
of disc growth and (10) whether disc particles are live or rigid.}
\label{tb:diskhalos}
\begin{center}
\begin{tabular}{l c c c c c c c c c}
\hline
(1)&(2)&(3)&(4)&(5)&(6)&(7)&(8)&(9)&(10)\\
name & halo & disc shape & $M_{d}$ &  $M_{h}$ & $\gamma$ & Q & major axes &
$t_{grow}$ & disc \\
\hline
1C&1&circular  &1&5&0.5&$\sim 1$& - &100&live\\
2C&2&circular  &1&5&0.5&$\sim 1$& - &100&live\\
3C&3&circular  &1&5&0.5&$\sim 1$& - &100&live\\
2E&2&elliptical&1&5&0.5&$\sim 1$&perpendicular&100&live\\
3E&3&elliptical&1&5&0.5&$\sim 1$&perpendicular&100&live\\
~\\
1'C&1&circular&1&5&5.0&$\sim 1$& - &100&live\\
2'C&2&circular&1&5&5.0&$\sim 1$& - &100	&live\\
3'C&3&circular&1&5&5.0&$\sim 1$& - &100	&live\\
2'E&2&elliptical&1&5&5.0&$\sim 1$&perpendicular&100&live\\
3'E&3&elliptical&1&5&5.0&$\sim 1$&perpendicular&100&live\\
~\\
1C t10&1&circular&1&5&0.5&$\sim 1$& - &10&live\\
2E t10&2&elliptical&1&5&0.5&$\sim 1$&perpendicular&10&live\\
3E t10&3&elliptical&1&5&0.5&$\sim 1$&perpendicular&10&live\\
1C t200&1&circular&1&5&0.5&$\sim 1$& - &200&live\\
2E t200&2&elliptical&1&5&0.5&$\sim 1$&perpendicular&200&live\\
3E t200&3&elliptical&1&5&0.5&$\sim 1$&perpendicular&200&live\\
~\\
1Cm&1&circular&	0.3&5&0.5&$\sim 1$& - &100&live\\
2Cm&2&circular&	0.3&5&0.5&$\sim 1$& - &100&live\\
3Cm&3&circular&	0.3&5&0.5&$\sim 1$& - &100&live\\
2Em&2&elliptical&0.3&5&0.5&$\sim 1$&perpendicular&100&live\\
3Em&3&elliptical&0.3&5&0.5&$\sim 1$&perpendicular&100&live\\
~\\
1C azi&1&circular&1&5&0.5&$\sim 1$& - &100&constrained\\
2C azi&2&circular&1&5&0.5&$\sim 1$& - &100&constrained\\
~\\
1C hot&1&circular&1&5&0.5&2.4& - &100&live\\
2E hot&2&elliptical&1&5&0.5&2.4&perpendicular&100&live\\
3E hot&3&elliptical&1&5&0.5&2.4&perpendicular&100&live\\
~\\
1C rigid&1&circular&1&5&0.5& - & - &100&rigid\\
2C rigid&2&circular&1&5&0.5& - & - &100&rigid\\
3C rigid&3&circular&1&5&0.5& - & - &100&rigid\\
~\\
3E 90&3&elliptical&1&5&0.5&$\sim 1$&parallel&100&live\\
\hline
\end{tabular}
\end{center}
\end{table*}

\begin{figure}
\includegraphics[width=\columnwidth]{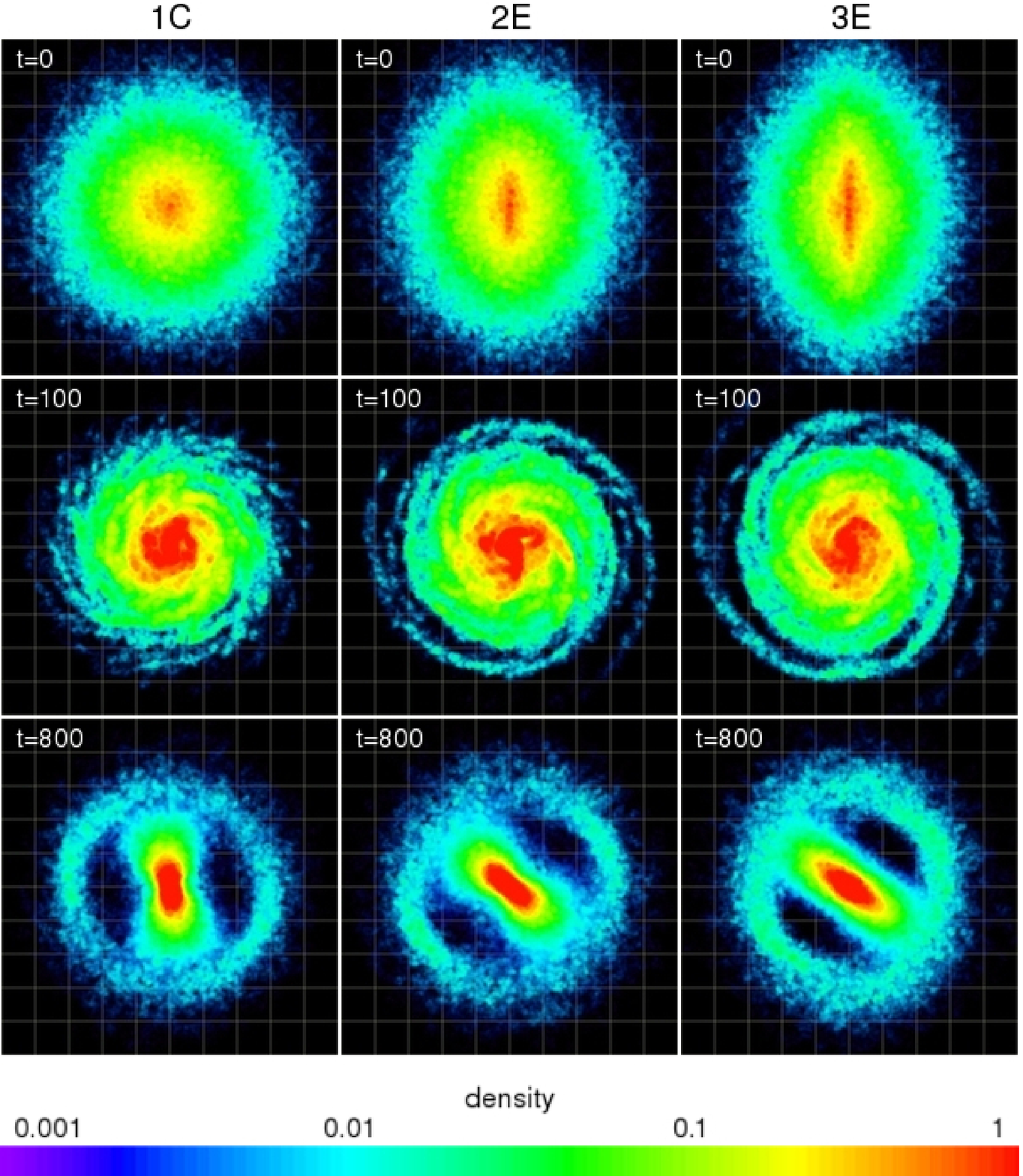}
\caption{Discs of models 1C (left), 2E (middle) and 3E (right) on the $xy$ plane
at $t=0$, $t=100$ and at $t=800$. Disc rotation is counterclockwise. Each
frame is 10 by 10 units of length. Colour represents projected density and the
range is the same for all panels. }
\label{fig_xy_disk}
\end{figure}

The triaxial halo initial conditions for our simulations 
were created using the iterative method of \citet*{Rodionov2009},
which creates equilibrium $N$-body systems with a given mass
distribution and, if desired, given kinematical constraints. In our
case we did not wish to impose specific kinematical constraints, so we
first adopted the desired mass distribution, as described below, and
then found the corresponding kinematics so that the model is in
equilibrium. For this we made a large number of successive
evolutionary steps of small duration, and at the end of each such step
brought back the mass distribution to the adopted density profile and
shape, as described in detail in \citet{Rodionov2009}. At the end of
this sequence we obtain initial conditions which are both in
equilibrium and have the desired mass distribution.  

The initially spherical haloes have the density profile described in
\citet{Hernquist1993}. They are then made triaxial by scaling the particle
positions in the $y$ and $z$ directions by factors of $b/a$ and $c/a$,
respectively ($c<b<a$). The iterative method is employed to obtain the
velocities of an equilibrium configuration with such shape. The density profile
of these triaxial haloes is described by:
\begin{equation}
\rho_{h}(r') = \frac{M_{h}}{2 \pi^{3/2}} \frac{\alpha}{r_{c}}
\frac{\exp{(-r'^{2}/r_{c}^{2})}}{r'^{2}+\gamma^{2}},
\end{equation}
where
\begin{equation}
r' = \sqrt{ (x/a)^2 + (y/b)^2 + (z/c)^2 },
\end{equation}
$M_h$ is the mass of the halo, $\gamma$ is a core radius and $r_{c}$ is the
cutoff radius.
The normalisation constant $\alpha$ is defined by
\begin{equation}
\alpha = \{ 1 - \sqrt{\pi} q \exp{(q^{2})} [1- \textrm{erf}(q)] \}^{-1}
\end{equation}
where $q=\gamma / r_{c}$ \citep{Hernquist1993}.

We employ in our simulations five haloes of different shapes: one
spherical and four triaxial ones, with the initial axis ratios given in Table
\ref{tb:halos}. All haloes have the same mass $M_{h}=5$ and cutoff radius
$r_{c}=10$. The core radius is $\gamma=0.5$ for all models (except in Section
\ref{sec:coresizes}, in which a different core size is explored). Haloes 1, 2
and
3 are used throughout the analysis, whereas the additional haloes 3-- and 3+ are
used mainly in Sect. \ref{sec:disklike}.

The haloes (with no disc) were evolved for 800 time units (units in Section
\ref{subsec:units}) to make sure that their shapes remain unchanged. Their axis
ratios are independent of radius at the beginning of the simulations and remain
so for 800 time units. The overall shape of the halo, taking all particles into
account, remains constant with time for all models. We may also measure the
shape using only particles in the inner region. In order to define this region,
we proceed as follows. We choose a radius, which represents the size of the
region we wish to study, in this particular case $r = 3$. We count the number
$n$ of particles inside a sphere of radius $r$. We sort all particles by density
and define the inner region as the region occupied by the first $n$ particles of
highest local density. In this way, we select an ellipsoidal shape (bounded by
an isodensity surface) and avoid delimiting the inner region by radius, which
would introduce a bias in the calculation of the shape
\citep{Athanassoula.Misiriotis02}. In the case of these pure haloes, however,
the inner shape does not differ from the overall shape, and both are constant in
time. Linear fits to the $b/a~(t)$ and $c/a~(t)$ of the pure halo models show
that the axis ratios typically change with a slope of the order of $10^{-6}$ to
$10^{-5}$. This means that in a Hubble time ($t\simeq1000$), the change in $b/a$
or $c/a$ is of the order of 0.1\% to 1\%. This holds for both the entire halo
and its inner region.

In the case of halo 3, we observe a vertical instability of the $m=4$ type.
When viewed on the $xz$ plane, halo 3 develops a transient X-like structure in
the beginning of the simulation. This instability is more pronounced if there is
a disc, but it is also measurable in the pure halo model. The relative intensity
of the $m=4$ Fourier component peaks at about $t=100$ but after that it strongly
subsides in both cases.

\subsection{Disc initial conditions}

\subsubsection{Epicyclic approximation to create elliptical discs}

If a circular disc were to be introduced in a triaxial halo, it would be out of
equilibrium. An elliptical disc whose shape is determined by the shape of the
halo potential should be initially closer to equilibrium and thus more suitable
as initial condition for the simulations. In order to set up the position and
velocity coordinates for such an elliptical disc, we use the epicyclic
approximation (\citealt{BinneyTremaine, GerhardVietri1986};
\citealt*{Franx+1994}). In
the presence of a 
non-axisymmetric halo potential, the disc particles are expected to have
non-circular orbits. This approximation tells us how elliptical each of these
orbits ought to be. Ultimately, the departure from circularity of each orbit is
determined by two quantities: the shape of the halo potential, but also its mass
distribution (in the form of circular velocity, $v_{c}$) -- both as a function
of radius. 

The first step is to create a circular disc with an exponential density profile:
\begin{equation} \label{rho_disk}
\rho_{d}(R,z) = \frac{M_{d}}{4 \pi z_{0} R_{d}^{2}}
\exp{\left(-\frac{R}{R_{d}}\right)}
\mathrm{sech}^{2}{\left(\frac{z}{z_{0}}\right)},
\end{equation}
where $M_{d}$ is the disc mass, $R_{d}=1$ is the scale length of the disc and
$z_{0}=0.2$ is the scale height. 

The elliptical disc will have an ellipticity of the orbits $\epsilon_{R}$ and an
ellipticity of the velocities $\epsilon_{v}$ (ellipticities are defined as
$\epsilon=1-b/a$), both of which have a radial dependence. Because the epicyclic
approximation does not take height into account, the vertical coordinates will
remain unchanged, and the $R$ and $\varphi$ coordinates of the disc will be
altered independently of $z$. The position coordinates on the plane are
reassigned as follows:
\begin{equation} \label{resumindo_pos_R}
R = R_{0} \left[ 1 - \frac{\epsilon_{R}}{2} \cos{(2\varphi_{0})} \right] \\
\end{equation}
\begin{equation} \label{resumindo_pos_phi}
\varphi = \varphi_{0} + \frac{\epsilon_{R}+\epsilon_{v}}{4}
\sin{(2\varphi_{0})},
\end{equation}
where $(R_{0},\varphi_{0})$ are the position coordinates of the particles of the
circular disc and $(R,\varphi)$ are the position coordinates of the particles of
the new elliptical disc. Similarly, the velocity coordinates will be:
\begin{equation} \label{resumindo_vel_R}
v_{R} = v_{c} \epsilon_{R} \sin{(2\varphi_{0})} \\
\end{equation}
\begin{equation} \label{resumindo_vel_phi}
v_{\varphi} = v_{c} \left[ 1 + \frac{\epsilon_{v}}{2} \cos{(2\varphi_{0})}
\right],
\end{equation}
where $v_{c}$ is the circular velocity, $v_{R}$ is the radial velocity and
$v_{\varphi}$ the tangential velocity. The ellipticities of the velocities, of
the positions and of the potential can be shown to have a simple dependence:
\begin{equation} \label{epsRvpot}
\epsilon_{v} = \epsilon_{R} +\epsilon_{pot}.
\end{equation}

Besides, it is possible to show that the ellipticity of the orbit,
$\epsilon_{R}$, is related to the ellipticity of the potential $\epsilon_{pot}$
through:
\begin{equation} \label{eps_Rpot}
\epsilon_{R} =  \epsilon_{pot} \left[ \left( \displaystyle
\frac{2v_{c}^{2}}{R}+\frac{dv_{c}^{2}}{dR} \right) \left( \displaystyle
\frac{2v_{c}^{2}}{R}-\frac{dv_{c}^{2}}{dR}  \right)^{-1} \right]_{R_{0}}, 
\end{equation}
which is a generalisation for any $v_{c}(R)$ of the particular case employed by
\citet{Franx+1994}, where the circular velocity was a power law.

So, for a given triaxial halo, we measure the ellipticity of the potential as a
function of radius. Then, measuring $v_{c}(R)$ and estimating its derivative
allows us to calculate the ellipticity of the orbits for each disc particle. The
particles in the disc have orbits whose ellipticities are not constant with
radius even if the ellipticity of the halo potential is. The epicyclic
approximation, however, is not valid when $v_{c}$ is approximately proportional
to $R$, and for this reason it can not be applied to the innermost part the
disc. To prevent it from diverging, $\epsilon_{R}$ is set to a constant value in
the innermost region.

The initial shapes of the halo potentials are quite independent of radius and
correspond to ($b/a$)$_{pot}$ of approximately 1, 0.85 and 0.72 for haloes 1, 2
and 3, respectively. The axis ratios of the potential are expected to be larger
than the axis ratios of the density because the isopotential contours (since
they refer to an integrated quantity) are always smoother and more circular than
the isodensity contours. In order to measure the axis ratios of the potential,
the halo particles are sorted by potential and the components of the inertia
tensor are calculated in consecutive intervals containing equal number of
particles. The ($b/a$)$_{pot}$ is obtained for each interval, interpolated to
the disc particle positions and used to calculate $\epsilon_{R}$ and
$\epsilon_{v}$ for each disc particle. The resulting shapes of the discs are
seen in the upper row of Fig. \ref{fig_xy_disk}. In equilibrium, the orbits of
the disc particles will be elongated in a direction perpendicular to that of the
halo major axis, and this comes out naturally from the epicycle approximation.
Besides creating elliptical discs, for comparisons, we also create equivalent
models using circular discs in each of the triaxial haloes (Table
\ref{tb:diskhalos}).

\begin{figure}
\includegraphics[width=\columnwidth]{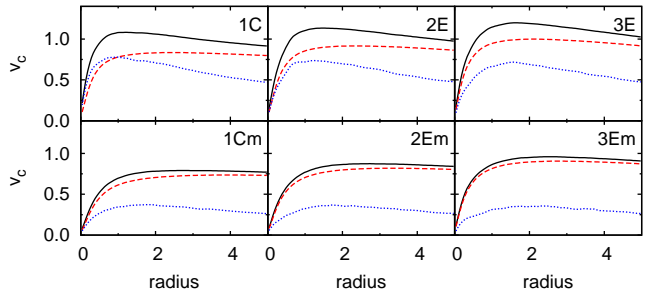}
\caption{Azimuthally averaged rotation curves at $t=100$: disc (dotted lines),
halo (dashed lines) and total (solid lines). Models with $M_{d}=1$ are shown in
the upper row and models with $M_{d}=0.3$ in lower row. In the upper right
corner of each panel we give the name of the corresponding model.}
\label{fig_vc_both}
\end{figure}

\subsubsection{Growing the disc}

Instead of introducing the disc abruptly, we slowly grow the mass of the disc
while the system evolves. This is done by gradually increasing the mass $m_{i}$
of each disc particle from (almost) zero to $M_{d}/N_{d}$, according to a smooth
curve:
\begin{equation}
m_{i}(t) = \frac{M_{d}}{N_{d}} \times \left\{
{\begin{array}{ll}
\displaystyle \frac{1}{2} \left( 1 - \cos{\frac{\pi
t}{t_{grow}}} \right)             &, ~0 \le t \le t_{grow}
\rule[-.5cm]{0cm}{1.cm}\\
\displaystyle 1 &, ~\qquad t > t_{grow}
\rule[-.5cm]{0cm}{1.cm}
\end{array}}\right.
\end{equation}
where $M_{d}$ is the final mass of the disc, $N_{d}$ is the number of disc
particles and $t_{grow}$ is the growth time of the disc mass. This procedure is
applied for an interval of 100 time units, during which time both the
halo and the disc are live. In the models described in Sect.
\ref{sec:timescales} other growth times are explored.

At first, the disc particles are very light and they feel the halo potential and
respond almost as test particles, without much self gravity and without
affecting the halo abruptly. At $t=100$ the disc is at full mass and, in some
respects, this is the instant that ought to be regarded as the actual beginning
of the simulation, since before this time both the total mass of the system and
its total angular momentum are in fact increasing. 

\begin{figure}
\includegraphics[width=\columnwidth]{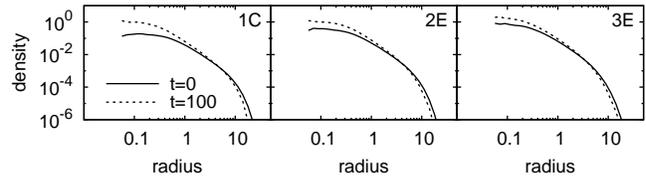}
\caption{Azimuthally averaged halo density profiles before ($t=0$, solid lines)
and after ($t=100$,
dotted lines) the introduction of the disc, for models 1C, 2E and 3E,
respectively.} 
\label{fig_haloprofiles}
\end{figure}

When the mass of the disc has finished growing ($t=100$) the contributions of
the disc and halo to the total circular velocity are comparable in the inner
region, for the spherical case (1C). In the more triaxial model (3E), however,
the halo is stronger even in the inner region. One of the reasons for this is
that the discs are not identical in the three models. The (azimuthally averaged)
rotation curve of the more elliptical disc (model 3E) is slightly lower in the
very centre, if compared to that of the circular disc in model 1C. Another
reason for the different rotation curves is the fact that the density of the
triaxial haloes is slightly higher in the centre if compared to the spherical
halo. As can be seen in Fig. \ref{fig_haloprofiles}, the density profiles at
$t=0$ (solid lines) are already slightly larger in the centre for the triaxial
haloes, because they are made triaxial by shifting the $y$ and $z$ coordinates
of the particles by factors smaller than one, which causes the concentration of
mass in the centre to increase. Apart from this fact, the inner halo density
suffers a further small increase (in all three models) due to the introduction
of the disc, presumably because the mass of the disc drags halo matter towards
the equatorial plane. This is why at $t=100$ (dotted lines in Fig.
\ref{fig_haloprofiles}) there has been a systematic increase of the density in
the innermost regions accompanied by a very mild but systematic density decline
for $r>10$. At $t=100$ the halo density profiles of the three models are roughly
similar, meaning that the initial differences in their inner densities were
compensated by the growth of the disc mass. So, as far as halo density profiles
are concerned, the three haloes are more similar at $t=100$ than they were at
$t=0$. For clarity, further times are not shown in Fig. \ref{fig_haloprofiles},
but from $t=100$ to $t=800$ the changes are insignificant.

Using the isothermal sheet approximation, the vertical component of the velocity
dispersion, $\sigma_{z}^{2}$, is determined by $z_{0}$:
\begin{equation}
\sigma_{z}^{2} = \pi G \Sigma(R) z_{0},
\end{equation}
where $\Sigma(R)$ is the surface density. The velocity dispersions in the other
two coordinates are acquired spontaneously by the disc during its evolution,
while its mass grows. Because of this method of growing the disc, where the
velocity dispersions are gradually acquired by the evolving disc, we do not set
a particular Toomre parameter $Q$ to begin with. By measuring the epicyclic
frequency $\kappa$ and the surface mass density $\Sigma$ we find that the values
of $Q$ for models 1C, 2E and 3E, averaged over the radius, are in the
range of 0.7 to 1.1, between $t=100$ and $t=140$. It should be noted that
the radially averaged values are meant as a rough estimate, since the radial
dependence of $Q$ is considerable. At later times the
bar grows and the potential becomes strongly non-axisymmetric, so that
the standard definition of $Q$ is not very meaningful.

\subsection{Miscellanea}
\label{subsec:units}

The units used here are such that the Newtonian gravitational constant
is $G=1$ and the scale length of the disc is $R_{d}=1$. Furthermore,
for the standard sequence of models $M_d=1$. In physical units, 
if we take the disc scale length to be 3.5 kpc and the mass of disc
 $5 \times 10^{10}$ M$_{\odot}$, for example, then the unit of time is $1.4
\times 10^{7}$ yr and the unit of velocity is 248 km/s.

The models here are evolved for 800 units of time, which, in the above example,
corresponds to 11.2 Gyr. In these simulations, the halo has $10^{6}$ particles
and the disc $2 \times 10^{5}$ particles.  The mass of the halo is always
$M_{h}=5$ and, for the standard models, the mass of the disc is $M_{d}=1$. The
evolution is calculated using the $N$-body code \textsf{gyrfalcON}
\citep{Dehnen2000, Dehnen2002}. In all cases, we used a softening of 0.05 units
of length, an opening angle of 0.6 and a time step of 1/64 units of time. This
led to an energy conservation of the order of 0.1\%.

\begin{figure}
\includegraphics[width=\columnwidth]{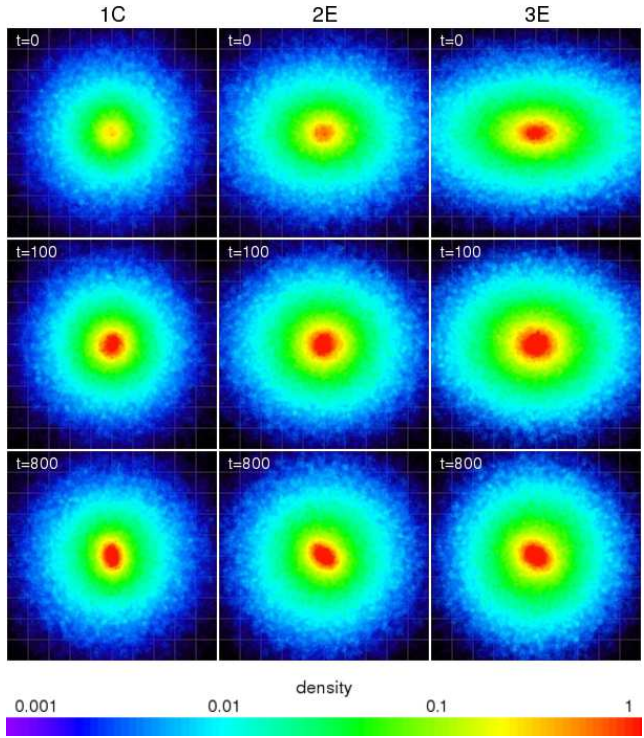}
\caption{Haloes of models 1C (left), 2E (middle) and 3E (right) on the $xy$
plane at $t=0$, $t=100$ and at $t=800$. Each frame is 10 by 10 units of
length. The projected density range is the same for all panels, and the same as
in Fig.~\ref{fig_xy_disk}. }
\label{fig_xy_halo}
\end{figure}

We calculate the $b/a$ and $c/a$ axis ratios using the eigenvalues of the
inertia tensor. If the shapes are measured in circular shells with equally
spaced radial bins, a strong bias towards sphericity is introduced. Instead, we
sort the halo particles as a function of local density and we measure the shape
inside density bins containing equal number of particles, as already
described in Sect. \ref{subsec:haloic}. These bins are not necessarily
spherical and they are not equally spaced in radius.

In order to measure the strength of the bars, we use the Fourier components of
the bidimensional mass distribution as a function of cylindrical radius,
computed in the following way:
\begin{equation}
a_m (R) = \sum _{i=0}^{N_{R}}~m_{i}~\cos (m\theta), ~ m=0, 1, 2, ...
\end{equation}
\begin{equation}
b_m (R) = \sum _{i=0}^{N_{R}}~m_{i}~\sin (m\theta), ~ m=1, 2, ... 
\end{equation}
where $N_{R}$ is the number of particles inside a given ring and $m_{i}$ is the
mass of each particle. The relative amplitude $A_{m}$ is defined as:
\begin{equation} \label{eq:Am}
A_m = \frac{\int_{0}^{R_{max}}~\sqrt{a_m^2+b_m^2}~R~dR}
{\int_{0}^{R_{max}}~a_0~R~dR  } .
\end{equation}
The `bar strength' is the quantity $A_m$ (for $m=2$), and the integration is
done until a maximum radius $R_{max}=3$, which is typically where the amplitude
of the $m=2$ component reaches a minimum.


\section{Loss of halo triaxiality due to elliptical or circular discs}
\label{sec:ECdisks}

\begin{figure}
\includegraphics[width=\columnwidth]{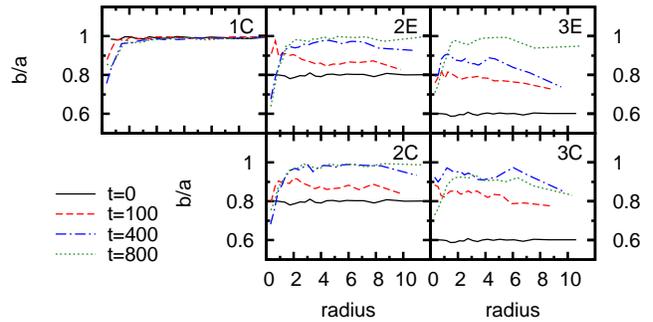}
\caption{The evolution of the $b/a$ radial profile. The first row shows the
evolution of models with haloes 1, 2 and 3 and with their respective equilibrium
discs. The second row shows the evolution of the same haloes, but with initially
circular discs.}
\label{fig_halosEC_ba_r}
\end{figure}

\begin{figure}
\includegraphics[scale=1]{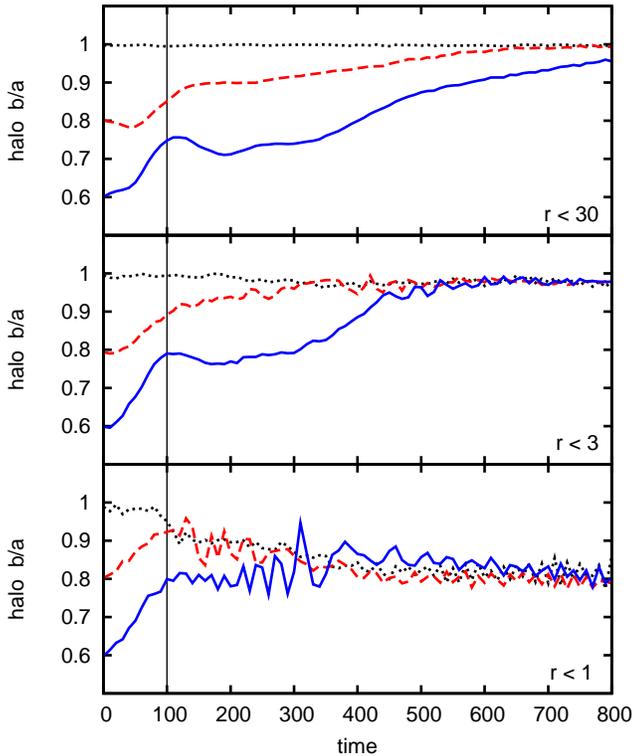}
\caption{The $b/a$ axis ratios of the halo, measured taking into account all
particles (top), the particles in $r<3$ (middle) and the particles in $r<1$
(bottom), as described in the text. The dotted, dashed and solid lines show
models 1C, 2E and 3E respectively.} 
\label{fig_ratios_alt4}
\end{figure}

\begin{figure}
\includegraphics[width=\columnwidth]{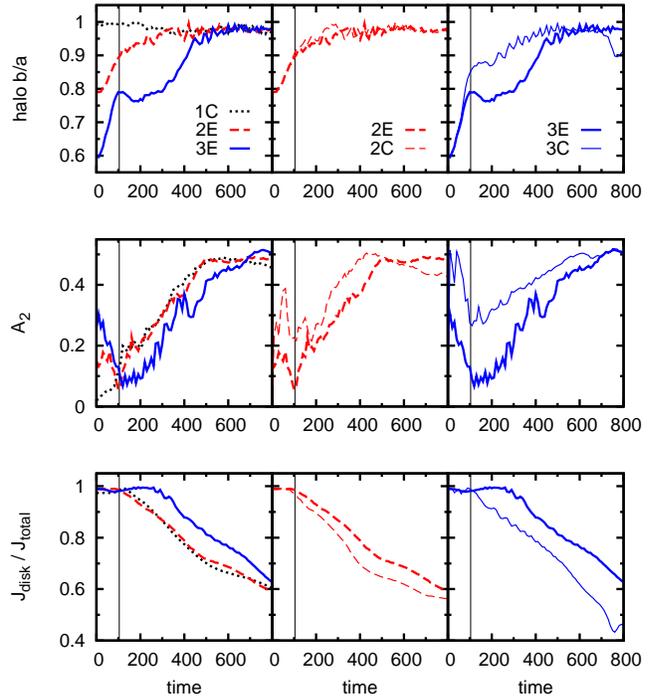} \\
\caption{Time evolution of inner halo $b/a$ ($r<3$), (top). $A_{2}$ (middle).
Fractional angular momentum of the disc, per unit mass (bottom). The first
column compares the spherical case with the two triaxial haloes containing their
respective elliptical discs. The dotted, dashed and solid lines show models 1C,
2E and 3E respectively. The second and third columns (for haloes 2 and 3
respectively) compare the elliptical disc (thick lines) and circular disc (thin
lines) cases.}
\label{fig_3babarmom}
\end{figure}

As described above we introduce the disc gradually over 100 time units and then
continue the fully self-consistent simulation to follow the evolution. The most
striking difference with respect to the pure halo models is that the haloes in
which (massive, bar-forming) discs are grown lose at least some of their
triaxiality (Fig. \ref{fig_xy_halo}). The effects on the vertical flattening
$c/a$ are presented in Sect. \ref{sec:halocovera}. Except for that section,
expressions such as `circularisation' and `loss of triaxiality' refer to the
face-on shape of the halo, i.e. they both mean that the halo becomes circular on
the disc equatorial plane of the disc ($b/a$ approaching 1), regardless of
$c/a$. 

\subsection{Standard models and models with initially circular discs}
\label{sec:EC}

In all models where (bar-forming) discs were grown, the haloes had their shapes
altered. Even in the case of a spherical halo with circular disc (1C) the
innermost region of the halo becomes rather prolate, due to the formation of the
bar: this is the ``halo bar'' \citep{Athanassoula2005a, Athanassoula2007}, or
``dark matter bar'' \citep*{Colin+2006}, which rotates together with the disc
bar, but is less elongated and less strong. The formation of such a structure
causes the halo $b/a$ to reach 0.8 in the inner region (Fig.
\ref{fig_halosEC_ba_r}) in all models where the disc forms a strong bar. Further
out, the initially triaxial haloes (models 2 and 3) lose their triaxiality
almost entirely after 800 time units. Roughly, almost half of the loss takes
place during disc growth (from solid to dashed lines in Fig.
\ref{fig_halosEC_ba_r}) and the second half takes place due to bar formation and
evolution (from dashed to dotted lines in Fig. \ref{fig_halosEC_ba_r}). This is
approximately valid also if circular discs are used instead of elliptical discs,
but there are interesting differences which will be discussed.

The time evolution of the shapes of models 1C, 2E and 3E are shown in Fig.
\ref{fig_ratios_alt4}. These are the standard models, because each of the three
haloes contains its equilibrium disc. The variations with circular discs (2C,
3C) will be used for comparisons. The halo $b/a$ are measured within three
different radii. First, they are measured taking all particles into account. To
measure the shape within a given radius $r$, we employ the procedure already
described in Sect. \ref{subsec:haloic}. This is done for $r=3$ and $r=1$.
Measuring the shapes in this fashion highlights respectively: the overall shape
(all particles), the shape at the region where most of the disc mass is located
($r<3$) and the shape in the region of the bar ($r<1$).

As can be seen in Fig. \ref{fig_ratios_alt4}, the shapes of the haloes in models
1C, 2E and 3E all tend to be same at $t=800$, the region of the disc ($r<3$)
approaching circularity faster than the outer halo. This is presumably due to
the fact that the dynamical time is much longer in the outer parts than in the
region with $r<3$. In $r<1$, all models develop the same halo bar with
$b/a=0.8$. The evolution of the vertical flattening (discussed in Sect.
\ref{sec:halocovera}) shows that even in the case of the spherical halo $c/a$
also drops to 0.8, which means that the halo bar is a prolate structure
(1:0.8:0.8, the circular plane containing the shorter axes).

\begin{figure}
\includegraphics[width=\columnwidth]{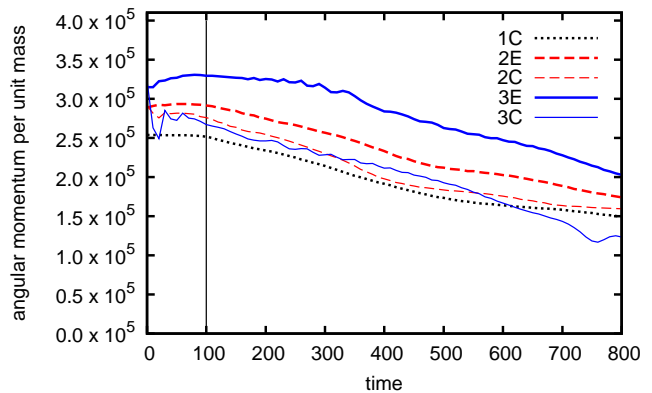}
\caption{The disc angular momentum is shown for models 1C (dotted), 2E (dashed)
and 3E (solid lines). Before $t=100$ the mass of the disc is still growing, but
the angular momentum per unit mass is roughly constant for most models. Also
shown is the angular momentum of models with circular discs (thin lines).}
\label{fig_angularmomentumAc}
\end{figure}

The comparisons of the central parts of models 1C, 2E and 3E (and also 2C and
3C) are shown in Fig. \ref{fig_3babarmom}. On the first row the halo $b/a$ for
$r<3$ is shown again for the three standard models and then separately for
models 2 and 3 comparing their respective elliptical and circular discs. The
more relevant difference is that for the case of the circular disc inside the
more triaxial halo (3C): the halo gets circularised faster than in the case of
the elliptical disc. During disc growth, the halo $b/a$ already increases more
for the circular disc and, from then on, it remains always larger than that of
model 3E, until about $t=500$. The second row of Fig. \ref{fig_3babarmom} shows
the evolution of the bar strengths. Bar strengths, given by the quantity
$A_2(R)$ (Eq. \ref{eq:Am}), are measured from $R=0$ to $R=3$, because that's
approximately the first minimum of $A_2(R)$ for all models with strong bars. For
the standard models, we note that at first, our measure of $A_{2}$ actually
decreases in the E cases. We point out that there is no actual bar at such
times: the non-zero $A_{2}$ of models E is simply a measure of the
non-circularity of the initial conditions, so that $A_{2}$ is in reality
indicating the non-axisymmetry, or non-circularity of the disc. During disc
growth this non-circularity is momentarily reduced, such that by $t=100$ all
three models have roughly the same $A_{2}$ value. Subsequently, the actual bar
instability sets in and a true bar grows in all three models, but with a delay
in the case of 3E. If we then compare what happens in the models with circular
discs and with elliptical discs, it turns out that, contrary to what one might
have naively expected, the bar is actually stronger in the \textit{initially}
circular discs. It should be emphasised, however, that these discs cease being
circular immediately after $t=0$, because they are presumably driven by the halo
into an excessively elliptical shape. At first, the $A_{2}$ values of the
circular discs increases very sharply, to values considerably higher than even
the initially elliptical discs. This behaviour is reminiscent of the transients
and overshoots observed in bar response calculations when the bar is not
inserted gradually. As a matter of fact, avoiding such unreasonable behaviour
was precisely one of the motivations for setting up initial conditions for
elliptical discs which should be in equilibrium. By $t=100$ the $A_{2}$ of the
circular discs has decreased somewhat, but it is still higher than that of the
elliptical discs. So by the time the bar begins to really form, it does so in a
disc which is actually more elliptical than in the E models. The result is that
in the C models, the bar is stronger than in the E models, during most of the
evolution.

The fact that discs with stronger bars circularise the halo more is a first
indication of the importance of bar formation and evolution in the loss of halo
triaxiality. In models that form stronger bars, the axisymmetrisation of the
halo is accomplished sooner. Comparing models 2E and 3E in Fig.
\ref{fig_3babarmom}, we see that during most of the time ($t=100$ to $t=400$ or
500) the bar in 2E is stronger than in 3E. At the same time, $b/a$ has larger
values. Similarly, in model 3, the E and C cases show that the bar of 3C is
stronger during $100<t<700$ and also the halo is clearly more circular. In fact,
the halo $b/a$ for 3E doesn't even increase very much in $100<t<400$. It only
becomes steeper at $t=400$, when the bar strength of 3E has finally caught up
with values comparable to those of 3C.

Another very important quantity in the evolution of these systems is the angular
momentum, whose total values (per unit mass) are shown in Fig.
\ref{fig_angularmomentumAc}. Each disc acquires a certain amount of angular
momentum while it is growing, and this amount is not the same for all models.
Thus the total angular momentum increases somewhat during a time equal to
$t_{grow}$, but stays constant after that. There is, however, considerable
angular momentum exchange between the disc and the halo component, as in models
with axisymmetric haloes (e.g. \citealt{Sellwood1980, DebattistaSellwood2000,
Athanassoula2003, ONeilDubinski2003}; \citealt*{Martinez-Valpuesta+2006};
\citealt*{VillaVargas+2009}). The net amount of angular momentum lost by
the disc is gained by the halo in all models. At $t=100$ the halo has nearly
zero angular momentum, so that roughly the total angular momentum of the system
is with the disc. Later, as the bar begins to form, it gets redistributed. One
important feature of these models is that the initially more elliptical discs
acquire more angular momentum. This can be clearly seen if we compare models 1C,
2E and 3E, our standard sequence of models with equilibrium discs (thick lines
in Fig. \ref{fig_angularmomentumAc}). And it can also be seen if we compare E
and C discs for one given halo: the initially circular discs (thin lines) have
less angular momentum than the equivalent initially elliptical disc in the same
halo, or, equivalently, in E models, the disc has acquired more angular momentum
by $t=100$ than in C models.

\begin{figure}
\includegraphics[width=\columnwidth]{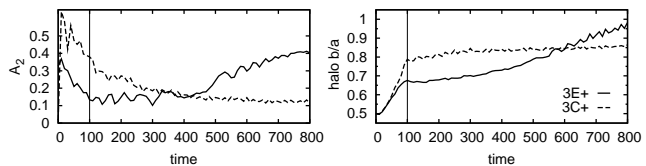} \\
\caption{$A_{2}$ and halo $b/a$ for models using the very triaxial halo 3+, with
circular disc (dashed lines) and elliptical disc (solid lines).}
\label{fig_dissolve}
\end{figure}

We may draw another similar observation from the lower panel of Fig.
\ref{fig_3babarmom}, which shows the fractional angular momentum of the discs.
We see that aside from having more total angular momentum, the initially more
elliptical disc will also hold a larger fraction of the total angular momentum
of the system than the fraction that will be held by an equivalent circular
disc, at any given time. This is valid for any one given halo. And it is still
true if we compare the elliptical discs of the two triaxial haloes. The
consequence of this is immediate: if a disc is holding on to its angular
momentum, it means that it is not forming a bar so well. So the discs that don't
lose their angular momentum so efficiently will have smaller bar strengths. This
is in good agreement with the results found for axisymmetric haloes, where a
tight correlation can be found between the bar strength and the angular momentum
gained by the halo and lost from the disc \citep{Athanassoula2003}.

So we might propose the following scenario: as they form, elliptical discs
acquire a larger amount of angular momentum. Also, they don't lose their angular
momentum so efficiently and this causes them to develop weaker bars, which take
more time to circularise their triaxial haloes.

One particularly interesting comparison between circular and elliptical discs is
provided by models using halo 3+, which is more triaxial than halo 3. Halo 3--,
being intermediate in shape between haloes 2 and 3, shows an evolution which is
merely intermediate between those two cases. In the case of the very triaxial
halo 3+, on the other hand, the difference between using a circular disc or an
elliptical disc is drastic. Figure \ref{fig_dissolve} shows the evolutions of
$A_{2}$ and of halo $b/a$ for models 3C+ and 3E+. For the model with a circular
disc (3C+), $A_{2}$ peaks strongly very early on ($t\sim10$) and then decays
gradually to small but non-zero values. For the model with an elliptical disc
(3E+), $A_{2}$ decreases at first and then starts growing in a manner similar to
the other E cases, albeit with a considerable delay. Model 3C+ is the only one
of our simulations in which halo triaxiality could be said to have inhibited bar
formation. That, however, appears to be due to the inadequacy of using a
circular disc as the initial conditions. When, instead, we use an elliptical
disc in the same halo, a strong bar does form.

The evolution of halo shapes in models 3C+ and 3E+ (Fig. \ref{fig_dissolve})
shows that disc introduction causes some halo circularisation in both cases. It
is also clear that in the case where there is bar formation (3E+), the halo
suffers further circularisation and the period of more intense loss of
triaxiality coincides with the period of faster bar growth. In the case where
there is no significant bar formation (3C+), the halo is able to remain triaxial
throughout. These results point to the contribution of the bar as one of the
factors causing loss of halo triaxiality.

These results are also indications of very important differences that
arise depending on 
whether one uses circular, and manifestly out of equilibrium discs, or
elliptical, and near equilibrium ones, when simulating bar formation within
triaxial haloes.

\subsection{Elliptical disc parallel to the halo major axis} 
\label{sec:major}

The equilibrium configuration for the system of an elliptical disc inside a
triaxial halo is such that the major axis of the disc is perpendicular to the
major axis of the halo. Therefore, in all our initial conditions the elliptical
disc is oriented in this way. We, nevertheless, also explored one simulation in
which the major axes of disc and halo are parallel, knowing that this would be
well off equilibrium. We took model 3E and turned by $90^{\circ}$ the elliptical
disc in the initial conditions (let us call it model 3E90). The result is that
the evolution of this model is very similar to the evolution of model 3C. That
is to say, in model 3E90 the bar is stronger than in model 3E. The angular
momentum transfer is more steep and the circularisation of the halo takes place
faster. As a matter of fact, the bar in model 3E90 is even slightly stronger
than in model 3C, and thus the angular momentum is lost by the disc even faster
and the halo $b/a$ consequently increases more rapidly. The disc in model 3E90
behaves much in the same manner as the disc of 3C in the beginning of the
simulation: it becomes excessively distorted in the direction perpendicular to
the halo major axis. This confirms the general trend that a stronger bar will
cause greater halo circularisation. It also indicates how strong the
effect of out-of-equilibrium initial conditions can be, thereby
stressing the importance of starting the simulation in near-equilibrium.

\subsection{Position angles} 
\label{sec:angles}

In order to study the relative orientations of the various elongated structures,
we distinguish between two regions of the disc and two regions of the halo,
namely the inner and outer parts of each. We therefore define the following four
components: the \textit{disc bar} ($0<R<3$), the \textit{halo bar} ($0<R<1$),
the \textit{outer disc} ($3<R<10$) and the \textit{outer halo} ($3<R<30$). Using
models 1C, 2E and 3E we then measure the position angles of each of these
components as a function of time, using the particles contained within those
radii. The angles are obtained from the Fourier analysis, and correspond to the
direction of the elongation of the $m=2$ component. The above definitions of the
disc bar region and of the halo bar region take into account the typical lengths
of these structures, that extend namely to about $R=3$ and $R=1$. The
definitions of the outer disc and outer halo are meant to give an estimate of
the direction of the overall orientation of such structures, without being
contaminated by the bars. The outermost parts of the halo retain some residual
triaxiality which, although small, is sufficient to allow a reasonable
determination of the direction of its major axis.

Once both are formed, the disc bar and the halo bar rotate together. The 
halo bar forms sooner in the spherical case, while in the
triaxial cases, the formation of the halo bar is somewhat delayed. Once they are
sufficiently strong, their position angles roughly coincide throughout the rest
of simulation.

The orientation of the outer halo is ill defined in the spherical case. In the
triaxial cases, the major axis turns very slowly. We find that it takes about
700 or 800 time units to turn by $90^{\circ}$. Such slow tumbling was also
obtained by \citet{HellerShlosmanAthanassoula2007a}, after the period of
collapse. This is also in good agreement with the results of
\citet{BailinSteinmetz2004}, who measured the figure rotation of haloes from
cosmological simulations and obtained pattern speeds with a log-normal
distribution centred at approximately $\Omega_{p} =
0.148~h~\mathrm{km~s^{-1}~kpc^{-1}}$. This means that a typical halo would
rotate roughly $90^{\circ}$ during a Hubble time.

The outer shape of the disc also bears some interesting relations to the bar. In
the beginning of the simulation the elliptical disc is perpendicular to the
major axis of the halo. At first, the position angle of the outer disc remains
in the same direction. Gradually, after both bars are formed, the overall shape
of the disc acquires figure rotation, with a pattern speed comparable to that of
the bar but out of phase with respect to it. The phase difference between the
disc bar and the outer disc is of $90^{\circ}$ at first, but then decreases.

 
\section{Different halo core sizes}
\label{sec:coresizes}

\begin{figure}
\includegraphics[width=\columnwidth]{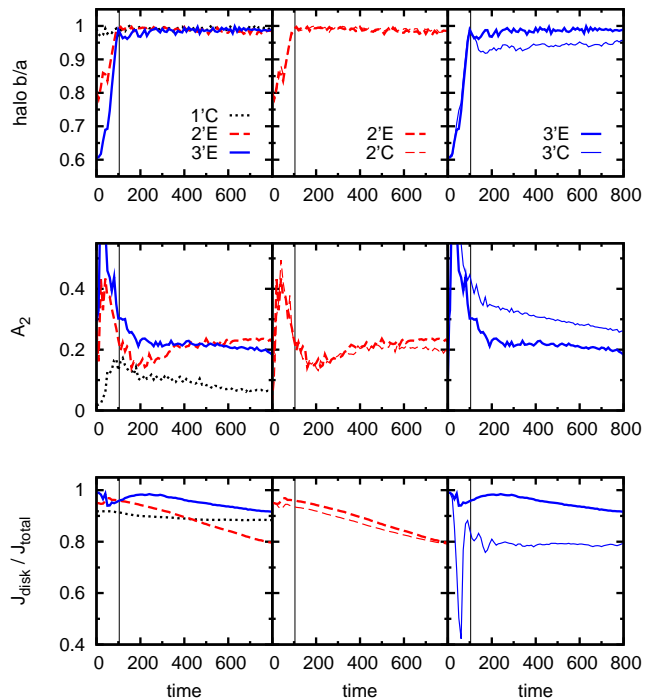} \\
\caption{Same as Fig. \ref{fig_3babarmom}, but for haloes with larger cores
($\gamma=5.0$). }
\label{fig_3babarmomEC50}
\end{figure}

In spherical haloes it was found that the core radius is an essential feature in
determining the evolution of the galaxy \citep{Athanassoula.Misiriotis02,
Athanassoula2003}. So far, we have only presented results from haloes with a
small core ($\gamma=0.5$), of the type called MH in
\citet{Athanassoula.Misiriotis02}. In this section we will discuss the change of
halo shape in simulations with large cores ($\gamma=5.0$, i.e. of the type
called MD in the aforementioned reference). These models, named 1'C, 2'E, 3'E
(and 2'C, 3'C), are the equivalent of the standard set 1C, 2E, 3E (and 2C, 3C)
in the sense that they have the same shapes as the standard set. However, they
have different density profiles, namely they are less concentrated. Evolved
without discs, these haloes with $\gamma=5.0$ are just as stable as the haloes
with $\gamma=0.5$, as far as their density profile and shape are concerned, i.e.
they retain their shapes and profiles until the end of the simulation.

Haloes with a large core have a peculiarity which causes the epicycle
approximation to give worse results. Because the core is larger, the velocity
curve is, at the start of the simulation, approximately linear ($v_{c} \propto
R$) across a larger region. The epicyclic approximation breaks down in this
case, which means that $\epsilon_{R}$  needs to be truncated at an arbitrary
value over a good portion of the disc. In the spherical case, the disc is
appropriately circular, but because of these difficulties, the disc shape is not
set quite properly in the triaxial models of larger core. As a consequence we
have some transients and overshoots: a disc which is not in equilibrium with its
triaxial halo will respond to the ellipticity of the potential it feels by
becoming excessively elongated at first, thus causing $A_{2}$ to increase too
much. This also means that with such haloes, the behaviour of the elliptical
discs are not much better than the circular ones in the same triaxial haloes.

The evolutions of halo $b/a$, $A_{2}$, and angular momentum for haloes with
large cores are shown in Fig. \ref{fig_3babarmomEC50}. They show clearly that
these haloes are more susceptible to circularisation. The disc growth alone is
capable of driving the halo $b/a$ to 1 by $t=100$ in all cases. These now
spherical, large-cored haloes don't receive much angular momentum from their
discs, and thus don't develop strong bars. The fact that the departures from
axisymmetry in models 2' and 3' are higher than in 1' is mostly due to the fact
that the discs in the initially triaxial haloes were driven to be excessively
elliptical, even though some of the non-axisymmetry in 2' can be attributed to
the formation of a weak bar, that can be discerned on the snapshots. Moreover,
there is some degree of angular momentum exchange.

The main result of this section is that less concentrated haloes are unable to
retain their triaxial shape. The growth of disc mass inside such large-cored
haloes is already enough to make them totally spherical. Furthermore, models 
with such haloes do not develop strong stellar bars, but in any case,
there wouldn't be any triaxiality left to be erased by the bar later on.


\section{Different time-scales for disc growth}
\label{sec:timescales}

\begin{figure}
\includegraphics[width=\columnwidth]{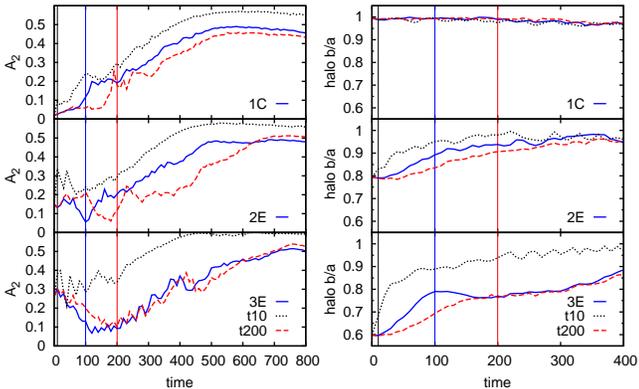}
\caption{$A_{2}$ and halo $b/a$ for simulations with different time-scales of
disc growth: $t_{grow}=10$ (solid lines), 100 (dotted lines) and 200 (dashed
lines). Note the different time scale in the right panels.}
\label{fig_tt}
\end{figure}

\citet{BerentzenShlosman2006} experimented with different ways of growing a seed
disc into a triaxial halo by adding the stellar particles gradually. They found
that the halo shape is not very sensitive to whether the disc is introduced
abruptly or quasi-adiabatically. Following their experiments, we also
re-ran models 1C, 2E, 3E both with a shorter ($t_{grow}=10$) and a longer
($t_{grow}=200$) time-scale of disc growth. For the first two models (top and
middle panels of Fig. \ref{fig_tt}), the result is that the overall evolution of
$A_{2}$ and halo $b/a$ is merely shifted to earlier or later times. We know that
the bar growth is faster in cases where the halo to disc mass ratio is smaller
\citep[e.g.][]{Athanassoula.Sellwood.1986}. So the temporal shift witnessed in
the two upper panels of Fig. \ref{fig_tt} could simply mean that disc mass has
to reach a certain limiting value for the bar to start growing sufficiently
rapidly. In models with smaller $t_{grow}$, the bar is comparatively stronger at
earlier times and the final value of $A_{2}$ is somewhat larger. 

In the case of model 3E, on the other hand, the evolutions of the bar strength
are rather similar in the cases where $t_{grow}=100$ and $t_{grow}=200$. The
$A_{2}$
usually begins to grow rapidly immediately after $t_{grow}$. In the
case of model 3E with the standard $t_{grow}=100$, however, it stalls for about
another 100 time units, and is eventually caught up with by model 3E with
$t_{grow}=200$. In model 3E ($t_{grow}=100$) the halo $b/a$ suffers a steep
increase during disc growth (going from 0.6 to 0.8). After that, this halo
momentarily undergoes a slight gain of triaxiality, during $100<t<200$. During
this period, instead of bar formation setting in, we have a small elongation of
the halo. By the time the halo has settled at $b/a\sim0.75$ it is
indistinguishable in shape from model 3E ($t_{grow}=200$). It is only then that
angular momentum transfer begins, for both models, and their bar strengths
increase simultaneously.


\section{Relative contributions of the disc and the bar to the loss
  of halo triaxiality} \label{sec:nobars}

The loss of halo triaxiality can be partly due to the introduction of the
disc and partly due to the growth of the bar. In order to disentangle these two
separate effects and to assess their relative contributions, we ran a number of
specifically designed simulations. This includes simulations with less massive
discs, simulations in which bar growth was artificially suppressed, simulations
with hot discs and simulations with rigid discs. 

\subsection{Less massive discs}
\label{sec:lessmassive}

\begin{figure}
\includegraphics[scale=1]{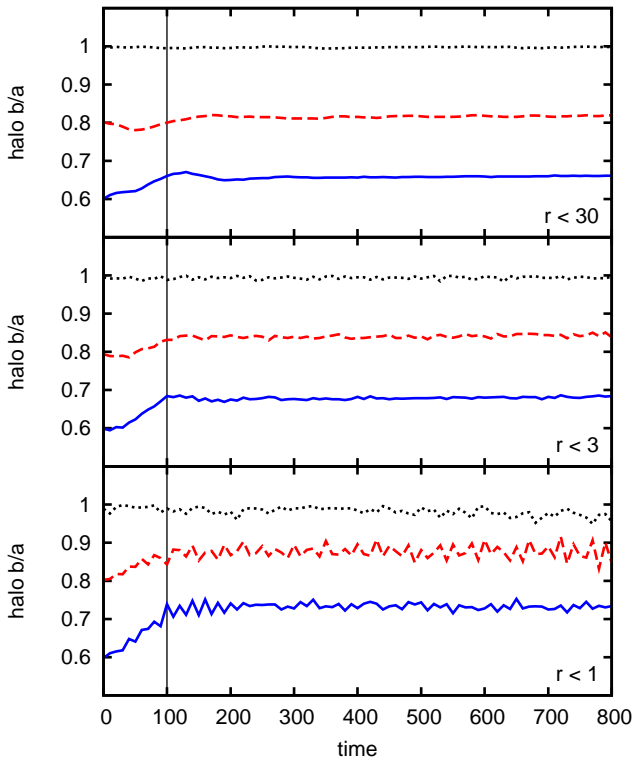}
\caption{Time evolution of the halo non-axisymmetry ($b/a$) for models with a
disc less massive than in our standard case ($M_{d}=0.3$, instead of 1). The
upper panel corresponds to the whole halo, the middle one to the inner halo and
the lower one to the innermost halo (see Section~\ref{sec:ECdisks} for
definitions). The dotted, dashed and solid lines show models 1Cm, 2Em and 3Em
respectively.} 
\label{fig_ratios_alt4mmC}
\end{figure}

As already discussed in \citet{Athanassoula2002}, the relative halo mass
influences the bar in two quite different ways. First the halo-to-disc mass
ratio influences the growth time of the bar, in the sense that relatively more
massive haloes (i.e. relatively less massive discs) lead to slower bar growths
\citep[e.g.][]{Athanassoula.Sellwood.1986}. Then during the bar evolution, the
halo helps the bar grow by absorbing at its resonances the angular momentum
emitted by the bar \citep{Athanassoula2002}. In fact the strongest bars form
when there is optimum balance between emitters and absorbers and this can
determine the location of corotation \citep{Athanassoula2003}.  

Thus, by adopting very low mass discs, we should obtain weak bars. For this
reason, we ran simulations with a disc mass $M_{d}=0.3$, i.e. less than a third
of the disc used in all previously discussed models, which have $M_{d}=1$. This
is one of the cases that can allow us to investigate whether it is the bar or
the presence of the disc itself that causes the halo to become axisymmetric; or
rather, to quantify the contributions of these effects. It is to be expected,
however, that non-massive discs should obviously have smaller effects on the
halo. The mass of these discs contributes little to the total circular velocity
curves (Fig. \ref{fig_vc_both}). Also, halo density profiles do not suffer much
increase in the inner region due to disc growth if $M_{d}=0.3$.

\begin{figure}
\includegraphics[width=\columnwidth]{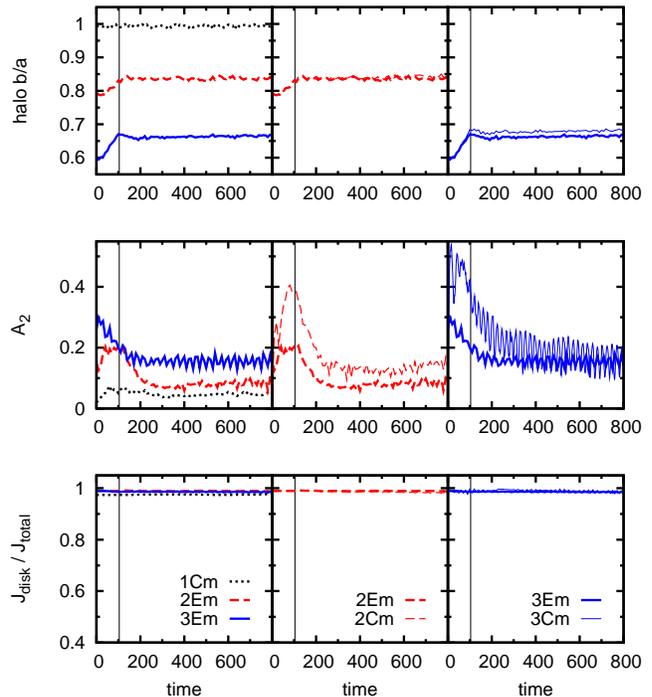} \\
\caption{Same as Fig. \ref{fig_3babarmom}, but for a less massive disc
  ($M_{d}=0.3$, instead of 1).}
\label{fig_3babarmomm}
\end{figure}

Figure \ref{fig_ratios_alt4mmC} shows the evolution of the halo $b/a$ for models
containing less massive discs, for the whole halo, the inner halo and the
innermost halo (see Section~\ref{sec:ECdisks} for definitions) and
Fig.~\ref{fig_3babarmomm} includes information on the disc non-axisymmetry and
the ratio of disc-to-total angular momentum. From the latter and from viewing
sequences of snapshots which show the evolution, we see that in all the models
with low disc mass there is no true bar formation, even though the inner disc
becomes elliptically distorted. Yet there is some, albeit little, change of the
halo shape, limited to $t<100$. Arguments from the time during which these
changes occur, the fact that there is no true bar and practically no angular
momentum exchange, lead to the conclusions  that the whatever loss of halo
triaxiality is witnessed is due to the introduction of the disc and not to any
subsequent bar formation.

And still we note that, during the period of disc growth, the halo $b/a$ in Fig.
\ref{fig_3babarmomm} does increase. In model 2Em it goes up to about 0.85 and in
model 3Em to little more than 0.65 (and slightly more so if the disc is
initially circular). This means that introducing a non-bar-forming disc caused
very small loss of halo triaxiality and only in the period of disc growth. 

Growing the more massive disc causes much greater loss of triaxiality. In Fig.
\ref{fig_3babarmom}, there is also an increase of halo $b/a$ due to disc
introduction, but it is much larger, specially in the case of the more triaxial
halo, where it grows from 0.6 to 0.8. This means that introducing a (massive)
bar-forming disc had caused a loss of triaxiality of about 0.2 in model 3E. This
is to be compared with a corresponding increase by 0.05 in model 3Em, which has
low mass disc. In the models with the standard disc mass there is further loss
of halo triaxiality after the disc has reached its full mass. Thus, by $t=800$
the halo is very near circular ($b/a>0.95$), which means that the amount by
which the halo shape changed during and after disc growth are comparable.
Equivalently in the case of the less triaxial halo 2E, the $b/a$ increases from
0.8 to 0.9 during disc growth and afterwards gets to about 0.95 as well. For the
models with lower mass discs we do not witness any further loss of halo
triaxiality after the disc has grown and the total change of triaxality is
rather small. 

To summarise, we can say the models with low mass discs suffer only a small loss
of halo triaxiality and that all of it is due to the introduction of the disc.
Although these models isolate the effect of the disc introduction, they can not
give us much information on the relative effects of the disc and bar for other
cases. They are, nevertheless, applicable to galaxies with low surface
brightness discs, and argue that such galaxies should not have suffered much
loss of halo triaxiality, and that their halo shape should be near what it was
from galaxy formation and as due to effects of interactions and mergings.

Before turning to other ways of assessing the relative role of the disc and bar
to the loss of halo triaxiality, we will discuss an interesting property
observed in one of our models, model 3Cm. Here we see that the $A_{2}$, the
strength of $m$ = 2 non-axisymmetry, shows oscillations. Viewing the evolution
of this model, we see that it did not develop a proper bar but has an elliptical
distortion in the centre (with some spiral structure) whose shape oscillates
periodically. This elliptical distortion rotates and its elongation is more
pronounced when it is perpendicular to the halo elongation. The $A_{2}$
amplitude always peaks when the orientation of the elliptical elongation is
perpendicular to the halo elongation. The oval rotates with a period of about 50
time units and even after 30 alignments, its mean strength has not decreased.

The behaviour of these discs is in some respects analogous to that of galaxies
with double-bars (also known as nested bars, or nuclear bars), in which there is
a primary (outer) bar and a secondary (inner) bar. In our simulations of
low-mass discs inside triaxial haloes, the discs don't develop bars, but the
oval distortion in the disc rotates in the presence of an elongated halo
potential. Thus the disc `oval' (to avoid calling it a bar) is analogous to a
`secondary bar' and the triaxial halo itself is analogous to a `primary bar',
with the difference that the triaxial halo does not rotate and its major axis
remains aligned with the $x$-axis.

In their theoretical approach to orbits within double bars, \citet{MS2000}, and
\citet{MaciejewskiAthanassoula2007} find that the loops supporting the inner bar
are thicker when the two bars are parallel; and that if the inner bar is a
self-consistent bar made of particles trapped around those loops, it should be
thinner when the two bars are perpendicular. \citet{DebattistaShen2007} describe
collisionless $N$-body simulations of discs in rigid haloes, which form double
bars. In such simulations, the bar strengths oscillate; the secondary bar
becomes stronger when the two bars are perpendicular and weaker when they are
parallel. Furthermore, the secondary bar rotates faster than average when they
are perpendicular; and slower than average when they are parallel. For our
analogous situation (which consists essentially of an elliptically distorted
disc rotating in an elongated halo potential) we obtain the same correlations,
which are also in agreement with the theoretical predictions of loops by
\citet{MS2000}. This would argue that low surface brightness galaxies whose
discs show important oval distortion could be living in haloes which are still
substantially non-axisymmetric.

\begin{figure}
\includegraphics[width=\columnwidth]{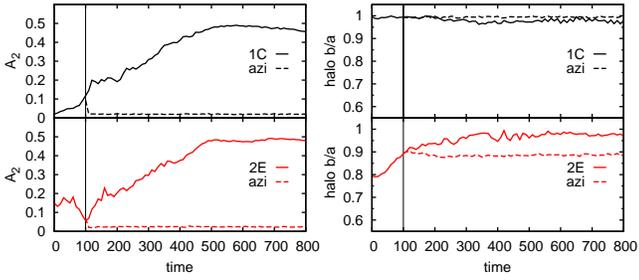}
\caption{$A_{2}$ and halo $b/a$ for models where bar formation has been
suppressed by imposing disc axisymmetry (dashed lines). The disc is
axisymmetrised at intervals of $\Delta t=1$. For comparison, the solid lines
show the results of the corresponding unconstrained simulations where
the bars do form.} 
\label{fig_aziB2}
\end{figure}

The $A_{2}$ amplitude is calculated with bidimensional quantities. In order to
have some estimate of the vertical shape of the oval, we compute $c/a$ for the
disc particles using the inertia tensor, in the same way as we do for halo
shapes. The vertical thickness of the disc (in cases 3Em and 3Cm) also
oscillates periodically. The oval is vertically thinner when it is perpendicular
to the halo elongation; and it is vertically thicker when it is parallel to the
halo elongation. It means that the oval's elongation correlates with its
vertical flattening: when the oval is more elongated, it is also more flattened.
Apart from oscillating, the mean thickness increases with time.

The haloes in simulations with low-mass discs remain triaxial. In models 3Em and
3Cm, the shape of the overall halo remains quite constant at about $b/a\sim0.7$,
with no significant radial dependence. In the innermost part of the halo
($r<1$), however, the halo $b/a$ oscillates, but by no more than 2\%. These
oscillations are very small, but measurable and quite regular. Furthermore, they
anti-correlate sharply with the shape of the oval: when the oval is more
elongated, the inner halo is less elongated. But this is not reminiscent of a
`halo bar', because there is no halo rotation at any radius: these haloes don't
rotate importantly (except for tumbling slowly), which means that the major axis
remains in the same direction.

\subsection{Suppressing bar formation by imposing disc axisymmetry}
\label{sec:randomazimuth}

In order to separate the effects of the bar and of the introduction of the disc,
we need to analyse simulations in which the disc is standard (i.e. not low
mass), but where there is no bar. We try to achieve this artificially, by
randomising the azimuthal coordinate of the disc particles in regular time
intervals ($\Delta t=1$) during the evolution of the system. Artificially
forcing the disc to retain axisymmetry throughout the evolution prevents any
non-axisymmetric structure from forming. We use the very same discs as in the
standard models, which would otherwise form strong bars, and observe the
behaviour of their respective haloes in the case where their bar formation is
suppressed by axisymmetrisation.

This technique consists in reassigning the $\varphi$ coordinate of the disc
particles to a random value between 0 and $2\pi$, while keeping their
cylindrical radius and their distance from the equatorial plane unaltered at
each intervention. This procedure is applied repeatedly during the evolution,
from the moment when the disc mass is fully grown, until the end of the
simulation and is straightforward in the case of axisymmetric haloes. In our
case, however, the potential of the halo is not axisymmetric, so that the
particles can find themselves in a region of deeper or shallower potential than
before the rotation. Because of that, we re-assign their velocities in such a
way as to conserve total energy, while at the same time keeping the angular
momentum of each particle unaltered. Tests showed that if the randomisations are
discontinued, then bar formation promptly sets in again.

\begin{figure}
\includegraphics[width=\columnwidth]{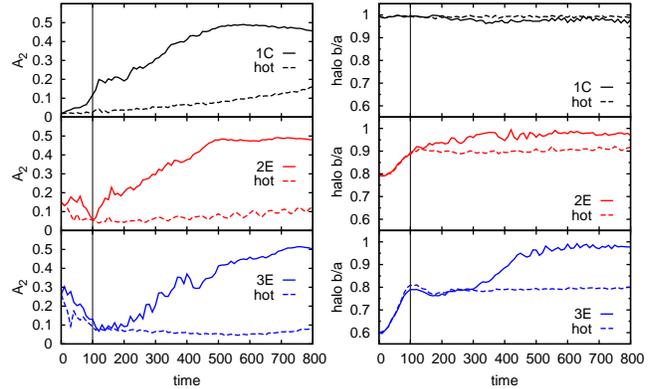}
\caption{Comparison between models with hot  discs ($Q=2.4$,
dashed lines) and models with cooler discs ($Q=1$, solid lines). The
left panels compare the $A_{2}$ and the right ones the halo $b/a$. 
}
\label{fig_hot73}
\end{figure}

This procedure works well for the spherical halo case because the discs in
such haloes are indeed meant to be circular (axisymmetric). In the case of the
triaxial haloes, making the disc perfectly axisymmetric causes it to be out of
equilibrium with the halo potential. In the case of halo model 2, bar formation
was successfully suppressed. However, in the case of halo model 3 it was not
possible to apply the randomisations without causing the disc to become severely
unstable. Experimentation showed that applying the interventions at different
intervals also caused the disc of model 3 to be disrupted by the end of the
simulation. Some disc particles end up gaining too much velocity and escape.
Since it was not possible to have a permanently circular and stable disc inside
a very triaxial halo, we exclude halo 3 from the analysis of this section.

Our purpose here is to evaluate the changes of halo shape in the absence of bar
formation. The evolutions of $A_{2}$ and $b/a$ are shown in Fig.
\ref{fig_aziB2}, for models 1C and 2E and for the corresponding simulations
where the axisymmetrisations described here were applied. The two evolutions
before $t=100$ are identical, because no axisymmetrisation was applied while the
disc grew, and, for the initially triaxial halo, we witness an increase of the
halo $b/a$ from 0.8 to 0.9.

After we start applying the axisymmetrisation, however, the evolutions become
different. In the unconstrained simulation, the bar forms and the triaxial halo
completely loses its remaining triaxiality and becomes quite axisymmetric by the
end of the simulation. In the simulations where the disc is continuously
axisymmetrised, however, the bar does not form and  there is no further loss of
halo triaxiality. So in the absence of bars, the halo of model 2E retains its
$b/a$ of 0.9, which otherwise would have gone to 1. 

The conclusion is that, in these models, a certain fraction (approximately half)
of the loss of halo triaxiality can be attributed to bar formation. It should,
however, be remembered that the axisymmetrisation process artificially keeps the
model out of equilibrium.

\begin{figure}
\centering
\includegraphics[width=0.49\columnwidth]{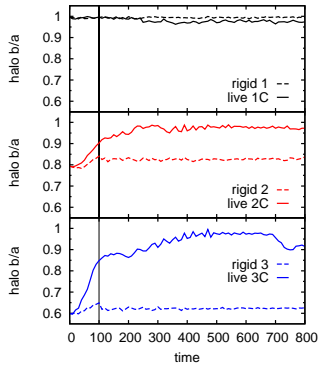}
\caption{Comparison of halo $b/a$ between models with a live circular
disc (solid lines) and the corresponding models with a rigid disc
(dashed lines).} 
\label{fig_rigid}
\end{figure}

\subsection{Hot discs}
\label{sec:hotdisks}

In axisymmetric haloes, hot discs are known to form oval distortions rather than
strong bars \citep{AthanassoulaIAU100,Athanassoula2003,Athanassoula2005a}. We
can thus use such discs to investigate whether they are able to completely
circularise their haloes, as bar-forming discs do. As initial conditions, we
first create circular discs (with Toomre parameter of $Q=2.4$), using the method
of \citet{Rodionov2009}. These discs are meant to be in equilibrium with a
spherical halo potential as that of halo 1. They are then made elliptical using
the epicyclic approximation, so that their shapes will be in equilibrium with
triaxial haloes 2 or 3. In these cases, however, only the positions are altered,
while the velocities remain those of a circular disc model, so as to keep the
velocity dispersions of the discs. Although this is not strictly correct, such
models are nevertheless closer to equilibrium than models of circular discs
inside triaxial haloes. During the period of disc mass growth, the velocities
gradually adapt to the elliptical potential, while not losing their higher
dispersions, which is important to this analysis.

In agreement with what was found in axisymmetric haloes, the models with hot
discs can reasonably be said not to have formed bars (Fig. \ref{fig_hot73}), as
$A_{2}$ remains always below 0.1 (except for model 1Chot where it
begins to grow a little towards the end of the simulation). The non-zero albeit
small values of $A_{2}$ are merely due to slight oval distortions in the disc.
These hot discs lose practically no angular momentum to their haloes (not shown
here). The consequence of the lack of bar formation is that again the halo is
able to remain triaxial (right panel of Fig. \ref{fig_hot73}). The $b/a$ of halo
2 goes from 0.8 to 0.9 and that of halo 3 goes from 0.6 to 0.8 in the period
between $t=0$ and $t=100$, for both the hot and the normal discs. After that,
however, the hot disc loses no further triaxiality, contrary to that of the
standard models discussed in Sect. \ref{sec:ECdisks}.

And so, as was the case also with the haloes of Sect. \ref{sec:randomazimuth},
there is no further circularisation besides that which was caused by the disc
growth. This is compelling evidence that indeed the bar plays an important role
in altering the shape of the halo.

\subsection{Rigid discs}
\label{sec:rigiddisks}

\begin{figure}
\includegraphics[width=\columnwidth]{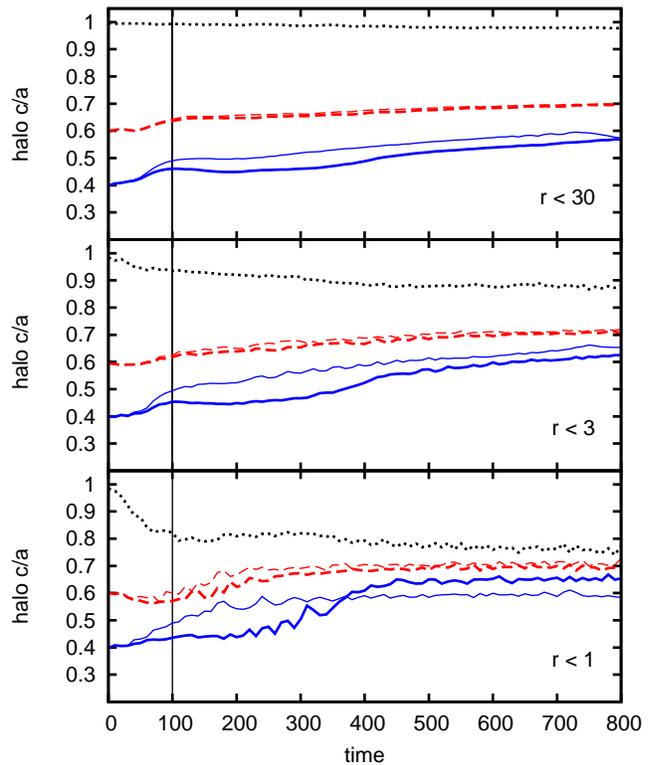}
\caption{Halo $c/a$ evolution for models 1C (dotted lines), 2E (dashed lines),
3E (solid lines), measured using particles within $r<1$ (bottom), $r<3$ (middle)
and all particles (top). Thin lines correspond to the respective models with
initially circular discs.}
\label{fig_73ca}
\end{figure}

Another way of evaluating the effects of bar-forming discs, as opposed to
non-bar-forming ones, is to replace the disc particles by an analytic potential.
In such a case, the simulation consists only of the usual halo particles, but
the disc is represented by a fixed potential, which is rigid and permanently
axisymmetric. The halo particles feel their own self gravity and they feel the
disc potential that has the same mass and scale lengths as in the simulations
with live discs before the disc instability sets in. As in previous cases, we
grew the potential gradually into the halo, according to a smooth function of
time, between $t=0$ and $t=100$.

The results in Fig. \ref{fig_rigid} compare the halo shape evolution of models
using live discs and of models with rigid discs. The latter suffer a small
amount of circularisation, only in the period of disc growth. The
circularisation induced by rigid discs is much smaller than that caused by live
discs, and it is in fact smaller even than that caused by low-mass live discs
(Sect. \ref{sec:lessmassive}). Once this rigid, circular, disc potential is in
place, there is no further change of halo shape. Since there is evidently no bar
formation, this again hints in the direction that the presence of the bar has
determining effects on the halo shape evolution. For comparison, the models with
live discs shown on Fig. \ref{fig_rigid} are the ones with circular discs, so
that the initial conditions are identical in the sense that the shapes and
masses of the disc are the same. The rigid disc simulations show us what happens
if such circular discs are forced to remain circular and not develop a bar.
Simulations with rigid potentials are, however, not realistic because there is
no exchange of angular momentum.

\begin{figure}
\includegraphics[width=\columnwidth]{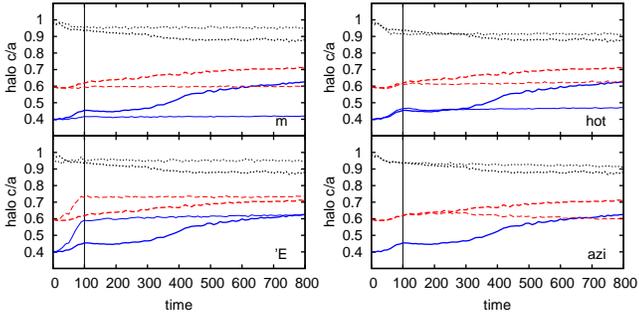}
\caption{Comparison of $c/a$ evolution between the standard models (thick lines:
1C, dotted;
2E, dashed; 3E, solid) and other models (thin lines). Four sets of models
are shown: less massive disc (upper left),
less concentrated halo (lower left), hotter disc (upper right) and
azimuthally randomised disc (lower right).}
\label{fig_canobarmC}
\end{figure}

\begin{figure}
\centering
\includegraphics[width=0.5\columnwidth]{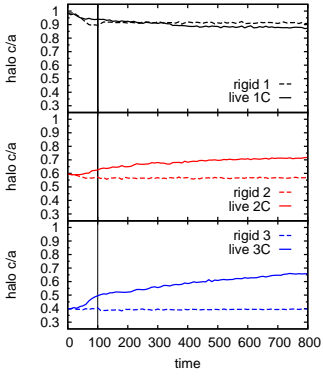}
\caption{Comparison of halo $c/a$ between models with a rigid disc (dashed
lines) and the corresponding models with a live circular disc (solid lines).}
\label{fig_rigidca}
\end{figure}

\section{Vertical shapes} \label{sec:covera}

\subsection{Halo vertical flattening} \label{sec:halocovera}

The minor to major axis ratio $c/a$ of the haloes is also affected both by disc
growth and by bar formation, but to a lesser degree than the intermediate to
major axis ratio $b/a$. The $c/a$ generally increases, indicating that the halo
tends to a less flattened configuration and this, taken together with the its
circularisation, shows that the halo tends to become more spherical. Models 1, 2
and 3 start out with axis ratios of 1:1:1, 1:0.8:0.6 and 1:0.6:0.4, respectively
and end with oblate shapes of roughly 1:1:0.9, 1:1:0.7 and 1:1:0.6 (as measured
in $r<3$). The only relevant radial dependence that is introduced in $c/a$ is
due to the presence of the disc, which makes the innermost regions somewhat more
flattened than the overall shape. In the case of the spherical model, the inner
region becomes significantly flatter and this change takes place mostly before
$t=100$. Presumably, due to the growth of disc mass, halo matter is pulled
towards the plane $z=0$ and parts of the halo in the immediate surroundings of
the disc become flatter than the overall shape. Figure \ref{fig_73ca} shows the
evolution of $c/a$ for the standard models 1C, 2E, 3E as measured within three
different radii.

\begin{figure}
\centering
\includegraphics[width=0.7\columnwidth]{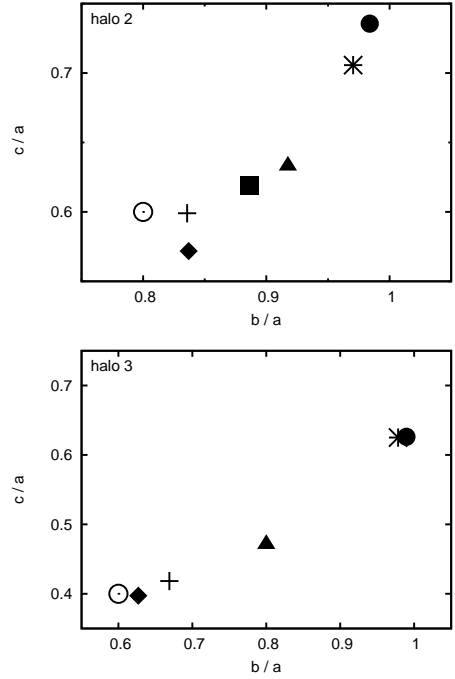}
\caption{Halo $c/a$ as a function of $b/a$ for models with halo 2 (top) and with
halo 3 (bottom). Note the different scales. The open circles mark the shapes of
the halo at $t=0$. The other symbols show the halo shape of each model at
$t=800$: standard models (asterisk), larger halo core (filled circle), less
massive disc (cross), axisymmetrised disc (square), hot disc (triangle) and
rigid disc (diamond).}
\label{fig_baca}
\end{figure}

In the simulations using the less massive disc, $c/a$ remains virtually
unaffected (Fig. \ref{fig_canobarmC}), showing not even a slight increase during
disc growth. These haloes remain approximately as triaxial as at $t=0$. The
models with a less concentrated halo suffer an increase of $c/a$ only until
$t=100$ (the time during which their $b/a$ goes to unity). After that, the $c/a$
does not change anymore and these haloes are oblate by the end of the simulation
(Fig. \ref{fig_canobarmC}).

In the other models where there is no bar formation (hot disc and axisymmetrised
disc (Fig. \ref{fig_canobarmC})), the $c/a$ increases only slightly until
$t=100$. The comparison of the non-bar-forming models with the standard models
shows that in the presence of bar formation the triaxial haloes would have
become still less flattened. This indicates that the bar acts not only on the
shape of the halo on the equatorial plane, but also affects its vertical
flattening; it causes the halo to tend towards sphericity by making it rounder
in both directions. In the models with no bar formation, the final shape of the
haloes is truly triaxial, being roughly 1:0.9:0.6 for halo 2 and 1:0.8:0.5 for
halo 3. Finally, in the spherical model, the halo $c/a$ suffers a small decrease
due to disc introduction, in both rigid disc and live disc cases. In the
triaxial models, live discs cause a small increase of $c/a$, but with the rigid
discs, there is hardly any change of $c/a$ (Fig. \ref{fig_rigidca}).

The final value of $c/a$ depends on the model, and the bar-forming models are
the ones in which $c/a$ changes the most. The range of variation of $c/a$ is
narrower than that of $b/a$, but clearly there is a correlation between the
amounts of $b/a$ and $c/a$ increase (Fig. \ref{fig_baca}).

\subsection{Formation of boxy/peanut bulges}
\label{sec:peanut}

\begin{figure}
\centering
\includegraphics[width=\columnwidth]{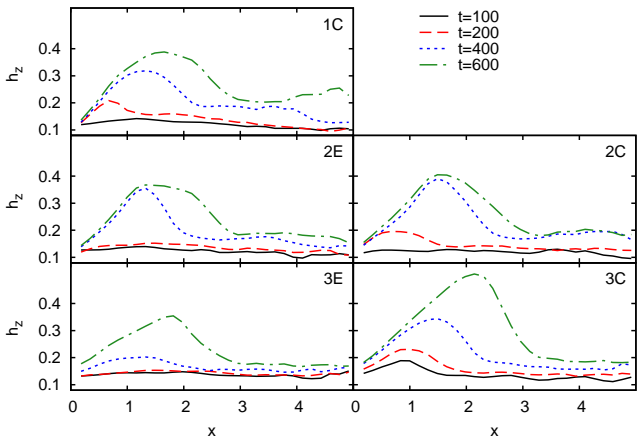} 
\caption{Dispersions of $z$ along the bar major axis.}
\label{fig_hzradius}
\end{figure}

\begin{figure}
\centering
\includegraphics[width=\columnwidth]{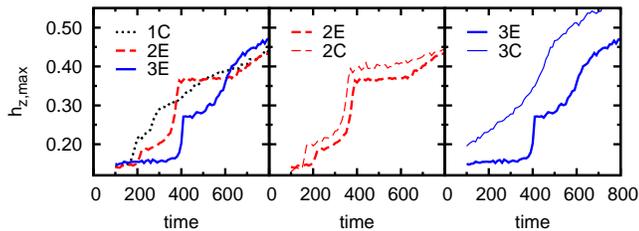} 
\caption{Strength of the peanut as a function of time.}
\label{fig_hztimes}
\end{figure}

In simulations with axisymmetric haloes, the discs of strongly barred galaxies
show a peanut-shaped structure, when viewed edge-on along the minor axis of the
bar. The peanut consists of two prominent humps that swell vertically from the
plane of the disc, on both sides. It begins to grow some time after the bar and
it becomes significantly stronger after the buckling, when the disc momentarily
loses its symmetry with respect to the $z=0$ plane \citep{CombesSanders1981,
Combes+1990, Raha1991, Athanassoula2005b, Athanassoula2008,
Martinez-Valpuesta+2006}.

Such structures were also observed in the simulations of
\citet{BerentzenShlosmanJogee2006} and \citet{BerentzenShlosman2006}, as well as
in ours, showing that halo triaxality does not inhibit their formation. We here
want to go one step further and check quantitatively the effect of triaxiality
on the peanut strength. In order to measure the latter quantity, we follow one
of the methods proposed by \citet{Athanassoula.Martinez-Valpuesta.2010}. We
first determine the orientation of the bar and then rotate the disc such that
the bar major axis lies in the direction of the $x$-axis. Considering then the
disc particles projected on the $xz$ plane, we measure the dispersion of the $z$
coordinates in successive slices of $\Delta x$. This dispersion -- denoted
$h_{z}$, to avoid the symbol normally used for velocity dispersions -- is an
indicator of the thickness of the peanut as a function of $x$. When the peanut
forms, $h_{z}$ reaches a maximum at a position $|x|$ which is near but within
the end of the bar, while remaining small in the centre (Fig.
\ref{fig_hzradius}). As the peanut becomes stronger, the maximum of $h_{z}$
increases, and the position of the maximum moves further out.

The evolution of peanut strength as a function of time, in simulations with
spherical haloes, has been described by \citet{Athanassoula2008}. Fig.
\ref{fig_hztimes} shows the $h_{z,max}$ as a function of time, for different
models. We find that in the triaxial models, the formation of the peanut is
delayed with respect to the spherical case. Furthermore, for any given halo
triaxiality, the C model forms the peanut sooner than the corresponding E model.
This is consistent with the fact that the peanut strength is related to the bar
strength \citep{Athanassoula.Martinez-Valpuesta.2010}. Since the C models tend
to develop stronger bars slightly earlier than the E models, it is expected that
they would also grow a strong peanut earlier.

We also measure the skewness $S_{z}$ of the distribution of vertical
coordinates, with respect to $z=0$, which is a measure of departures from
vertical symmetry: high values of $S_{z}$ correspond to a buckling of the disc.
Each sharp increase of the peanut strength $h_{z,max}$ coincides with a buckling
episode (not shown here). The time of the first buckling increases with the
triaxiality of the model. And a C model buckles before the corresponding E
model.


\section{Kinematics of the disc-like halo particles}
\label{sec:disklike}

\citet{Athanassoula2007} showed that in strongly barred galaxies, the inner
parts of the halo display some mean rotation in the same sense as the disc
rotation. This is more important for particles nearer to the equatorial plane
and decreases with increasing distance from it, but is always much smaller than
the disc rotation. Here we extend this analysis to triaxial haloes and point out
some kinematic properties that depend on the initial triaxiality of the haloes.

If we select halo particles in a region around the equatorial plane ($|z|<0.5$)
and measure their tangential velocities $v_{\varphi}$ at a given time ($t=800$)
we already notice that there is some rotation, with peak velocities of about
$v_{\varphi}=0.1$. This definition, however, may include particles which
happened to be passing by the equatorial region a $t=800$, but that are not
permanently staying close to it. We, therefore, use two alternative definitions
to select these disc-like halo particles: we select the particles that are
within $|z|<0.5$ at $t=800$, but that have remained inside this cylinder during
$800<t<900$ (definition 1) or during $600<t<1200$ (definition 2). The first
definition already removes many of the particles that were not truly rotating,
but definition 2 is even more strict.

\begin{figure}
\centering
\includegraphics[scale=0.9]{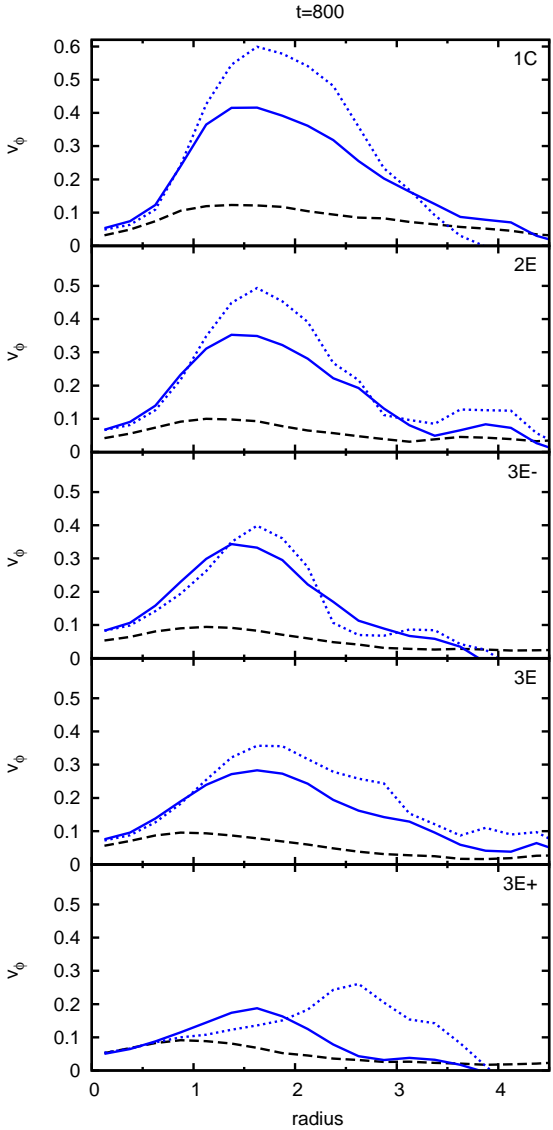}
\caption{Radial profile of tangential velocities, measured at $t = 800$, for the
permanently disc-like halo particles, using definitions 1 (solid lines) and 2
(dotted lines), as discussed in the text. The corresponding profiles for the
$|z|<0.5$ region are also shown (dashed lines).}
\label{fig_vels_keep_800}
\end{figure}

The radial profiles of tangential velocities are shown in Fig.
\ref{fig_vels_keep_800}, for particles simply within $|z|<0.5$ at $t=800$ and
also for the two other definitions. Note that the tangential velocities are
significantly higher when the more strict requirements are applied to define the
region of disc-like particles; they range from $v_{\varphi}=0.2$ to 0.6,
depending on the model and the definition. The spherical halo shows more
rotation than the triaxial ones and indeed the peak tangential velocities
decrease with increasing triaxiality. More precisely, it should be stated that
the tangential velocity depends on the \textit{initial} triaxiality of the
haloes, because by $t=800$, these five haloes have approximately the same
shapes. And yet their kinematics at that time still depend systematically on the
initial shape.

\begin{figure}
\includegraphics[scale=1]{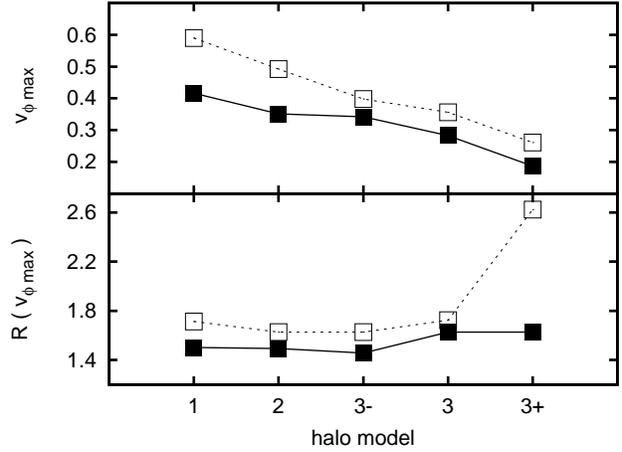}
\caption{Top: Peak tangential velocities, measured at $t = 800$, of
  the disc-like halo particles using 
definitions 1 (solid squares) and 2 (open squares). Bottom: Radii of the peak
tangential velocities.}
\label{fig_peaks}
\end{figure}

The peak tangential velocities of the five models are shown in the upper panel
of Fig. \ref{fig_peaks}. In models whose halo was initially more triaxial, there
is less rotation. The lower panel of Fig. \ref{fig_peaks} shows the radii at
which the tangential velocities peak.

\begin{figure}
\includegraphics[width=\columnwidth]{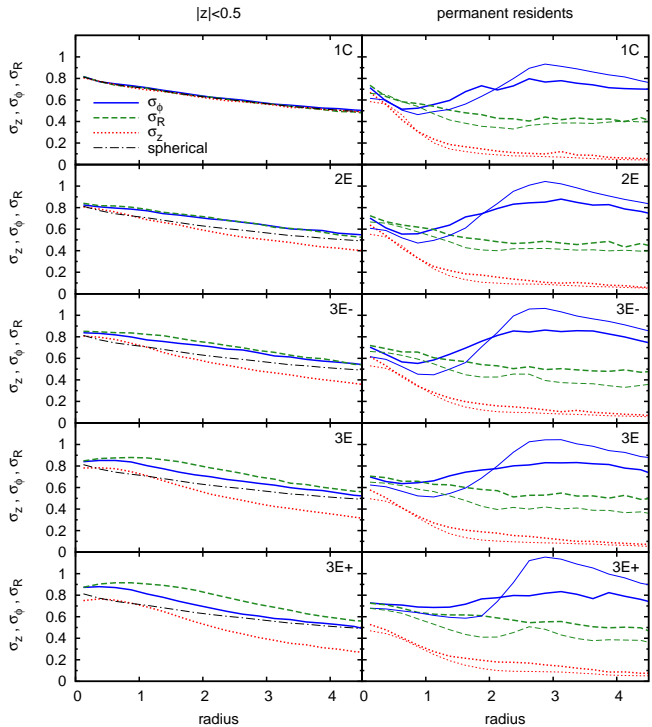}
\caption{Left: velocity dispersions $\sigma_{\varphi}$ (solid lines),
$\sigma_{R}$ (dashed lines) and $\sigma_{z}$ (dotted lines) for the $|z|<0.5$
region. For comparison, the average velocity dispersion of the spherical case is
show in all panels (dot-dashed line). Right: velocity dispersions for the
permanently disc-like halo particles, by definitions 1 (thick lines) and 2 (thin
lines), as discussed in the text. The line types are as for the left panels. All
panels, both left and right, correspond to $t = 800$.} 
\label{fig_sigs_keep_800}
\end{figure}

The velocity dispersions of the spherical halo are isotropic. The initially
triaxial haloes retain an anisotropy, even if by $t=800$ they also have become
spherical. The left panel of Fig. \ref{fig_sigs_keep_800} shows the radial
profiles of the tangential, radial and vertical velocity dispersions for the
$|z|<0.5$ region of the five haloes. The more triaxial haloes have
systematically larger radial dispersions and systematically smaller vertical
dispersions (the centre excluded). Such feature one would expect to be due to
construction of the triaxial haloes, but they are still present at $t=800$ when
all haloes have become roughly spherical. Additionally, it can be noticed that
the departures from isotropy increase with initial triaxiality of the model.
Again we note that although such anisotropy  in the velocity dispersions would
be obvious in the initial conditions, it is not evident that it would be
retained after the haloes have been circularised and have all reached roughly
the same shapes. In the right panel of Fig. \ref{fig_sigs_keep_800}, the
velocity dispersions are shown for the two definitions of the region with
disc-like kinematics. In this case, as one would expect for particles that are
rotating, the vertical velocities are much smaller and the tangential velocities
are much higher than the isotropic velocity dispersions of the spherical case
(and slightly more so with definition 2 than with definition 1). With these two
definitions, there is no systematic dependence of anisotropy on initial shape
(except perhaps in the innermost region, where the spherical halo is more
isotropic). But generally, using definitions 1 and 2, the disc-like halo
particles show the same regime of rotation for all five models, with the
anisotropy peaking at about $r=3$. Note also that the difference between the
results from definitions 1 and 2 increases with increasing initial triaxiality.
If the entire halo is taken in account, the velocity dispersions are similar to
those of the left panel of Fig. \ref{fig_sigs_keep_800}.

\begin{figure}
\centering
\includegraphics[scale=1]{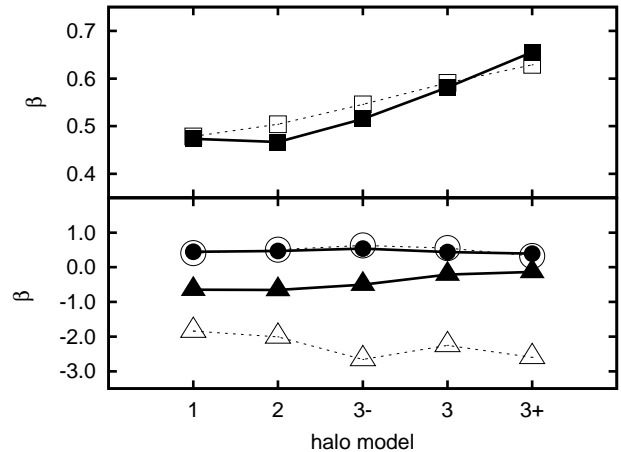}
\caption{Top: anisotropy parameter for the entire halo (filled squares) and for
the $|z|<0.5$ region (open squares). Bottom: anisotropy parameter with
definitions 1 (solid symbols) and 2 (open symbols), measured at $R=1$ (circles)
and at $R=3$ (triangles). Note that the two panels have very different scales.}
\label{fig_betas}
\end{figure}

In order to quantify the anisotropy, we use an anisotropy parameter $\beta$
defined as:
\begin{equation} \beta = 1 - \frac{1}{2} \left(
\frac{\sigma_{\varphi}}{\sigma_{R}} \right)^{2}.
\end{equation}
In the case of isotropy, $\beta$ would be 0.5. The value of $\beta$, calculated
from the whole halo, increases with increasing initial triaxiality, which means
that the radial motions are correspondingly more important, even at $t=800$
(upper panel of Fig. \ref{fig_betas}). If only particles within $|z|<0.5$ are
taken into account, the anisotropy is similar (upper panel of Fig.
\ref{fig_betas}). When using the entire halo or the $|z|<0.5$ particles, $\beta$
does not have much radial dependence and does not show important changes with
time. The lower panel of Fig. \ref{fig_betas} shows the anisotropy for the
disc-like halo particles using definitions 1 (solid symbols) and 2 (open
symbols). With these definitions and when the anisotropy is measured at $R=1$
(circles), there is some isotropy in the very centre, since there $\beta$ is
close to 0.5. When $\beta$ is measured at its peaks ($R=3$), the anisotropy is
larger and shows more important tangential motions (note that the two panels
have very different scales). With the more strict definition 2, the anisotropy
is even higher. But with either definition and at any radii, there is not a
significant dependence of anisotropy (of the disc-like halo particles) with halo
model.


\section{Summary and conclusions}
\label{sec:conclusions}

Cosmological $N$-body simulations have shown that dark matter haloes of galaxies
should be triaxial, at least in cases where there are no baryons
\citep{Allgood2006}. This can be the result of asymmetric mergings, or of a
radial orbit instability \citep{Bellovary2008}, coupled to tidal effects from
other galaxies or from groups and clusters. The resulting prolate halo has very
little or no figure rotation.

The aim of this paper was to investigate how such a halo will influence bar
formation and, more generally, how it will influence the secular evolution of
disc galaxies. Our combined disc and halo initial conditions were built so as to
be as near equilibrium as possible, with discs which are initially elliptical.
We have, nevertheless, also considered initially circular discs, to test what
the effect of such less realistic initial conditions would be.

The growth and evolution of such discs drive the haloes rapidly towards
axisymmetry, except for the innermost parts, where the final shape of the halo
is elongated. This latter effect is independent of the initial halo triaxiality
and is found also in initially circular haloes \citep[e.g.][]{Athanassoula2005a,
Athanassoula2007, Colin+2006}. It is linked to the angular momentum exchange
within the galaxy and the formation of a `halo bar', which is shorter and less
elongated than the disc bar, but rotates with the same pattern speed.

This innermost prolate elongation put aside, the remaining halo tends towards
axisymmetry even for models with considerable initial triaxiality, and this from
the moment the disc starts growing. One can distinguish two different
axisymmetrisation phases. Initially, while the disc grows, this trend towards
halo axisymmetry is quite rapid. The second phase depends on whether or not a
bar is formed: in the presence of a bar, the circularisation continues.

The disc shape also changes with time. Initially it is elongated, its
ellipticity depending on that of the halo. Thus the $A_{2}$ is non-zero at $t =
0$, although there is no bar. It decreases with time and reaches a minimum
around the time that the disc has reached its maximum mass. After that, the bar
starts forming and induces a further increase of halo axisymmetry. In our
models, bar formation takes place in the presence of haloes that are still
considerably triaxial, even though the triaxiality is later erased. In the more
triaxial cases, bar formation is somewhat delayed, but the subsequent evolution
of the bar proceeds in a manner similar to the spherical halo case. We have also
presented triaxial models that do not develop bars at all; but in such cases,
the equivalent spherical ones do not either. The general agreement of our
results with previous studies is in the sense that a truly triaxial halo cannot
coexist with a strong bar for very long: one of these non-axisymmetries must
give way. We have presented simulations in which the bar prevails and the halo
triaxiality yields. This argues that in situations where the parameters are such
that a bar is known to form in the spherical case, it would also form in the
(initially) triaxial cases, further erasing the triaxiality as it does.

Using circular discs, instead of very near-equilibrium elliptical ones, may give
rise to quite different $A_{2}$ evolutions, particularly if the triaxiality of
the halo is important. In the case of a very triaxial halo containing a circular
disc, the $A_{2}$ initially increases very abruptly to reach a strong peak and
then decays. That was the only simulation in which halo triaxiality damped bar
formation, but we stress that this is only when a circular disc was used. In
that very triaxial halo, a circular disc is even further from equilibrium than
in the less triaxial ones. There are other situations in which $A_{2}$ grows
very abruptly in the very beginning and then decays without forming a bar, and
this only happens if we use circular discs as initial conditions. For instance,
in the case of low mass discs, the initial $A_{2}$ increase is due to the
circular disc becoming excessively distorted. After the peak, the remaining
non-zero $A_{2}$ is due to a mild oval distortion in the centre of the disc (and
also some vague spiral structure), but none of this amounts to actual bar
formation (furthermore, there is no exchange of angular momentum between disc
and halo). 

Such behaviour of $A_{2}$ might be analogous to those of
\citet{BerentzenShlosmanJogee2006}, who also have initially circular discs and
find that the bars in their live triaxial haloes dissolve after a few Gyr
(except in the case of their more cuspy halo, which does not retain its
triaxiality). Similarly, \citet{BerentzenShlosman2006}, also with
initially circular discs, find that bar formation is damped
by the triaxiality of the halo. In their models, strong bars form in discs that
have erased the halo prolateness almost completely.

In all our models, as in previously run models with initially spherical haloes,
bar formation is always necessarily accompanied by angular momentum exchange;
more specifically, angular momentum is transferred from the disc to the halo. In
the beginning of the simulations, practically all the angular momentum is in the
disc. In all models where there is bar formation, the disc-to-total ratio of
angular momentum decreases steadily. Further properties of the angular momentum
exchange are similar to those found in spherical haloes; e.g. it is related to
the bar strength; strong bars causing more transfer from disc to halo. Also, in
haloes with large core, the bars are weaker and the angular momentum lost by the
disc is correspondingly small. There are simulations where the $A_{2}$ is
non-zero due to oval distortions of the disc which, however, do not qualify as
real bars. In such cases, there is very little, if any, angular momentum
exchange between disc and halo. We also find differences in angular momentum
transfer depending on whether we use circular or elliptical discs as initial
conditions. The initially circular discs become excessively elongated during the
very short phase of disc growth and they thus develop strong bars faster then
the equivalent model with an elliptical disc. These discs, that host stronger
bars, lose angular momentum faster. The other quantity that is also closely
related to angular momentum transfer and bar strength is the shape of the halo:
a stronger bar causes the halo to become axisymmetric faster.
 
Our standard models have haloes with a small core, of the type called MH in
\citet{Athanassoula.Misiriotis02}. Since the core radius is an essential feature
in determining the evolution of the galaxy \citep{Athanassoula.Misiriotis02,
Athanassoula2003}, we also ran simulations with large cores, i.e. of the MD
type. Here the effect of the disc is even stronger. In all models, by the time
the disc was fully grown, the halo has already become nearly axisymmetric, so
that any further evolution follows closely that of the MD models of
\citet{Athanassoula.Misiriotis02}, namely, compared to MH models, the bar is
less strong and there is less angular momentum exchange between the disc and the
halo.

We also presented models in which the growth time of the disc was either much
smaller, or much bigger than that of the standard models. We found that this
does not affect much the results, limiting itself mostly to a temporal shift of
the evolution, backwards or forwards in time, respectively. This is in
good agreement with similar tests made by \citet{BerentzenShlosman2006},
albeit with different models.

A crucial question in this context is how much of the loss of the halo
triaxiality is due to the introduction of the disc and how much to the formation
of the halo. In the case of MD discs, i.e. of discs which are heavy relative to
the halo in the inner parts of the galaxy, we saw that by the time the disc has
reached its full mass, the halo has become nearly spherical. It is thus
reasonable to expect that low mass discs would have a smaller effect on the halo
shape, as already found in the simulations of \citet{BerentzenShlosman2006}. We
repeated such simulations here, but with initially elliptical discs. In these
cases, our disc has sufficiently low mass so that very little angular momentum
can be emitted from its inner regions. Thus although the halo was ready to
receive angular momentum, not much exchange was possible for lack of sufficient
emitters. As a result, the bar was exceedingly weak, as in the very low mass
discs of \citet{Athanassoula2003}, and so brought no effect on the halo
triaxiality. Thus in these cases the slight decrease of the halo ellipticity
when the disc is introduced, is not followed by further decrease caused by the
bar. We point out that our corresponding spherical halo model with low mass disc
does not develop a bar either, showing that it was not the triaxiality of the
other models that was responsible for inhibiting bar formation. In fact, in none
of our sets of models with initially near-equilibrium discs, do we find cases
where a bar develops in the spherical halo, while failing to do so in the
equivalent triaxial ones. Such models, with discs of low mass relative to the
halo, give us useful information on galaxies with low surface brightness discs,
and argue that such galaxies should not have suffered much loss of halo
triaxiality, and that their halo shape should be near what it was from galaxy
formation and as due to effects of interactions and mergings.

We made further simulations to find what fraction of the triaxiality loss can be
attributed to the disc introduction and what to the formation and evolution of
the bar. For this we ran models where the disc was kept artificially
axisymmetric, models where the disc was very hot (so that bar grew very slowly
and was more of an oval than a strong bar) and models where the disc was rigid.
In these three cases (as well as in the low mass case), the simulations were
designed so as not to form bars. In all cases where there is no bar, the halo is
able to remain triaxial until the end of the simulation. Any circularisation
that takes place is restricted to the period of disc growth. This
circularisation is particularly small if the disc mass is small or if the disc
is rigid. The rigid discs -- where the potential of a (permanently circular)
disc is represented by an analytic potential instead of live particles -- are
the ones that cause the smallest effects on their haloes, suggesting that it is
not merely the presence of disc mass that alters the halo shape, but also the
active response of the disc orbits that oppose the halo elongation. 

We find that the vertical shape of the triaxial haloes is also affected both by
disc growth and by bar formation, in the sense that haloes become less flattened
with time. These changes are, however, always small compared to the changes in
the equatorial shape. Nevertheless, in cases where the $b/a$ increase is more
pronounced, that of $c/a$ is important also. The general evolution of vertical
shapes is thus analogous to that of the halo shape on the plane of the disc:
disc introduction causes some degree of change and the subsequent formation of a
bar determines whether or not there will be further loss of flattening. But even
in models where a strong bar forms (and the halo is thus completely
axisymmetrised), the vertical flattening remains considerable, meaning that the
haloes have become approximately oblate by the end of the simulation.

As the bar forms, the vertical structure of the disc is also affected. As with
previous simulations of axisymmetric galaxies, the discs in the triaxial models
also undergo buckling episodes and peanut formation. In our initially triaxial
models that form strong bars, the first buckling of the disc occurs later than
in the spherical case. Consequently, peanut formation is delayed in the
triaxial models. The models with initially circular discs develop stronger
peanuts before the corresponding models with elliptical discs. All this is in
agreement with the fact that peanut strength is related to bar strength
\citep{Athanassoula.Martinez-Valpuesta.2010}, and the fact that bars
grow faster in initially circular discs.

The spherical halo particles in a layer near the equatorial plane acquire during
the evolution some rotation in the same sense as the disc rotation, as in
\citet{Athanassoula2007}. We extended that analysis to triaxial haloes. We
select halo particles that are permanently in the vicinity of the $z=0$ plane;
by analysing their radial profiles of tangential velocities, we find that these
`disc-like' halo particles show considerable rotation, with velocities of the
order of half of that of the disc particles. Such rotation is present as well in
the triaxial models and the peak tangential velocities depend systematically on
the initial triaxiality of the model. There is more rotation in the spherical
halo than in the triaxial ones, even though their shapes are roughly the same by
the end of the simulation. Apart from the rotation, there is another kinematic
feature which is dependent on the initial triaxiality: the anisotropy of the
velocity dispersions. If we consider the entire haloes, we see that the in
spherical one the velocities are isotropic, but the initially triaxial models
retain their anisotropy even after their shapes have become approximately
spherical. We find that the velocity anisotropy at the end of the simulation
depends systematically on the \textit{initial} triaxiality.

To summarise, we have presented simulations of bar formation in triaxial haloes
and showed that the haloes become more axisymmetric due to two separate factors:
disc growth and bar formation. Typically half the circularisation can be
attributed to the introduction of the disc and the other half to the formation
of a strong bar. Halo vertical flattening is also affected, but to a much lesser
degree, meaning that haloes become roughly oblate by the end of the simulation.
This is the first study of bar formation and disc bar evolution in which live
elliptical discs within live triaxial haloes were employed as the initial
conditions for $N$-body simulations. Elliptical discs have the advantage of
being closer to equilibrium and of not responding to the presence of the
aspherical halo potential by becoming excessively and unphysically distorted, as
circular discs tend to do. We have also analysed the kinematics of the halo and
pointed out that even after the haloes lose their triaxiality, they are able to
retain the anisotropy of their velocity dispersions. We also note that the
disc-like particles of the halo rotate less in the haloes that were initially
triaxial.

Interesting issues remain to be explored. Simulations such as these, in which
all components are live, can be used to study in detail the orbital structure,
not only during the period of disc growth, but also during the period of bar
formation. Another interesting issue is the effect of gas on the dynamics of
these systems. Simulations analogous to the ones we have presented, but
including also a gas component are currently under way and will be the subject
of a forthcoming paper.


\section*{Acknowledgements}

We wish to thank the referee for his fast reply and careful
reading. REGM acknowledges support from the Brazilian agencies FAPESP
(05/04005-0) and CAPES (3981/07-0), and from the French Ministry of Foreign and
European Affairs (bourse Eiffel). This work was partly supported by grant
ANR-06-BLAN-0172.

\bibliographystyle{mn2e.bst}
\bibliography{myreferences.bib}

\begin{thebibliography}{}

\bibitem[\protect\citeauthoryear{{Allgood}, {Flores}, {Primack}, {Kravtsov},
  {Wechsler}, {Faltenbacher} \& {Bullock}}{{Allgood}
  et~al.}{2006}]{Allgood2006}
{Allgood} B.,  {Flores} R.~A.,  {Primack} J.~R.,  {Kravtsov} A.~V.,  {Wechsler}
  R.~H.,  {Faltenbacher} A.,    {Bullock} J.~S.,  2006, \mnras, 367, 1781

\bibitem[\protect\citeauthoryear{{Athanassoula}}{{Athanassoula}}{1983}]{Athana%
ssoulaIAU100}
{Athanassoula} E.,  ed. 1983, {in \textit{Internal Kinematics and Dynamics of
  Galaxies}, IAU Symp. 100, Dordrecht: Reidel, 432pp}

\bibitem[\protect\citeauthoryear{{Athanassoula}}{{Athanassoula}}{2002}]{Athana%
ssoula2002}
{Athanassoula} E.,  2002, \apjl, 569, L83

\bibitem[\protect\citeauthoryear{{Athanassoula}}{{Athanassoula}}{2003}]{Athana%
ssoula2003}
{Athanassoula} E.,  2003, \mnras, 341, 1179

\bibitem[\protect\citeauthoryear{{Athanassoula}}{{Athanassoula}}{2005a}]{Athan%
assoula2005a}
{Athanassoula} E.,  2005a, Celestial Mechanics and Dynamical Astronomy, 91, 9

\bibitem[\protect\citeauthoryear{{Athanassoula}}{{Athanassoula}}{2005b}]{Athan%
assoula2005b}
{Athanassoula} E.,  2005b, \mnras, 358, 1477

\bibitem[\protect\citeauthoryear{{Athanassoula}}{{Athanassoula}}{2007}]{Athana%
ssoula2007}
{Athanassoula} E.,  2007, \mnras, 377, 1569

\bibitem[\protect\citeauthoryear{{Athanassoula}}{{Athanassoula}}{2008}]{Athana%
ssoula2008}
{Athanassoula} E.,  2008, in \textit{Formation and Evolution of Galaxy Bulges},
  IAU Symp. 245, eds M. Bureau, E. Athanassoula and B. Barbuy, Cambridge
  University Press, 93

\bibitem[\protect\citeauthoryear{{Athanassoula} \&
  {Martinez-Valpuesta}}{{Athanassoula} \&
  {Martinez-Valpuesta}}{2010}]{Athanassoula.Martinez-Valpuesta.2010}
{Athanassoula} E.,  {Martinez-Valpuesta} I.,  2010, \mnras, in press

\bibitem[\protect\citeauthoryear{{Athanassoula} \& {Misiriotis}}{{Athanassoula}
  \& {Misiriotis}}{2002}]{Athanassoula.Misiriotis02}
{Athanassoula} E.,  {Misiriotis} A.,  2002, \mnras, 330, 35

\bibitem[\protect\citeauthoryear{{Athanassoula} \& {Sellwood}}{{Athanassoula}
  \& {Sellwood}}{1986}]{Athanassoula.Sellwood.1986}
{Athanassoula} E.,  {Sellwood} J.~A.,  1986, \mnras, 221, 213

\bibitem[\protect\citeauthoryear{{Bailin}, {Simon}, {Bolatto}, {Gibson} \&
  {Power}}{{Bailin} et~al.}{2007}]{Bailin2007}
{Bailin} J.,  {Simon} J.~D.,  {Bolatto} A.~D.,  {Gibson} B.~K.,    {Power} C.,
  2007, \apj, 667, 191

\bibitem[\protect\citeauthoryear{{Bailin} \& {Steinmetz}}{{Bailin} \&
  {Steinmetz}}{2004}]{BailinSteinmetz2004}
{Bailin} J.,  {Steinmetz} M.,  2004, \apj, 616, 27

\bibitem[\protect\citeauthoryear{{Bailin} \& {Steinmetz}}{{Bailin} \&
  {Steinmetz}}{2005}]{BailinSteinmetz2005}
{Bailin} J.,  {Steinmetz} M.,  2005, \apj, 627, 647

\bibitem[\protect\citeauthoryear{{Bellovary}, {Dalcanton}, {Babul}, {Quinn},
  {Maas}, {Austin}, {Williams} \& {Barnes}}{{Bellovary}
  et~al.}{2008}]{Bellovary2008}
{Bellovary} J.~M.,  {Dalcanton} J.~J.,  {Babul} A.,  {Quinn} T.~R.,  {Maas}
  R.~W.,  {Austin} C.~G.,  {Williams} L.~L.~R.,    {Barnes} E.~I.,  2008, \apj,
  685, 739

\bibitem[\protect\citeauthoryear{{Berentzen} \& {Shlosman}}{{Berentzen} \&
  {Shlosman}}{2006}]{BerentzenShlosman2006}
{Berentzen} I.,  {Shlosman} I.,  2006, \apj, 648, 807

\bibitem[\protect\citeauthoryear{{Berentzen}, {Shlosman} \&
  {Jogee}}{{Berentzen} et~al.}{2006}]{BerentzenShlosmanJogee2006}
{Berentzen} I.,  {Shlosman} I.,    {Jogee} S.,  2006, \apj, 637, 582

\bibitem[\protect\citeauthoryear{{Binney} \& {Tremaine}}{{Binney} \&
  {Tremaine}}{1987}]{BinneyTremaine}
{Binney} J.,  {Tremaine} S.,  1987, {Galactic Dynamics}.
Princeton University Press

\bibitem[\protect\citeauthoryear{{Cole} \& {Lacey}}{{Cole} \&
  {Lacey}}{1996}]{Cole1996}
{Cole} S.,  {Lacey} C.,  1996, \mnras, 281, 716

\bibitem[\protect\citeauthoryear{{Col{\'{\i}}n}, {Valenzuela} \&
  {Klypin}}{{Col{\'{\i}}n} et~al.}{2006}]{Colin+2006}
{Col{\'{\i}}n} P.,  {Valenzuela} O.,    {Klypin} A.,  2006, \apj, 644, 687

\bibitem[\protect\citeauthoryear{{Combes}, {Debbasch}, {Friedli} \&
  {Pfenniger}}{{Combes} et~al.}{1990}]{Combes+1990}
{Combes} F.,  {Debbasch} F.,  {Friedli} D.,    {Pfenniger} D.,  1990, \aap,
  233, 82

\bibitem[\protect\citeauthoryear{{Combes} \& {Sanders}}{{Combes} \&
  {Sanders}}{1981}]{CombesSanders1981}
{Combes} F.,  {Sanders} R.~H.,  1981, \aap, 96, 164

\bibitem[\protect\citeauthoryear{{Curir}, {Mazzei} \& {Murante}}{{Curir}
  et~al.}{2006}]{Curir+2006}
{Curir} A.,  {Mazzei} P.,    {Murante} G.,  2006, \aap, 447, 453

\bibitem[\protect\citeauthoryear{{Debattista}, {Moore}, {Quinn}, {Kazantzidis},
  {Maas}, {Mayer}, {Read} \& {Stadel}}{{Debattista}
  et~al.}{2008}]{Debattista2008}
{Debattista} V.~P.,  {Moore} B.,  {Quinn} T.,  {Kazantzidis} S.,  {Maas} R.,
  {Mayer} L.,  {Read} J.,    {Stadel} J.,  2008, \apj, 681, 1076

\bibitem[\protect\citeauthoryear{{Debattista} \& {Sellwood}}{{Debattista} \&
  {Sellwood}}{2000}]{DebattistaSellwood2000}
{Debattista} V.~P.,  {Sellwood} J.~A.,  2000, \apj, 543, 704

\bibitem[\protect\citeauthoryear{{Debattista} \& {Shen}}{{Debattista} \&
  {Shen}}{2007}]{DebattistaShen2007}
{Debattista} V.~P.,  {Shen} J.,  2007, \apjl, 654, L127

\bibitem[\protect\citeauthoryear{{Dehnen}}{{Dehnen}}{2000}]{Dehnen2000}
{Dehnen} W.,  2000, \apjl, 536, L39

\bibitem[\protect\citeauthoryear{{Dehnen}}{{Dehnen}}{2002}]{Dehnen2002}
{Dehnen} W.,  2002, Journal of Computational Physics, 179, 27

\bibitem[\protect\citeauthoryear{{Dubinski} \& {Carlberg}}{{Dubinski} \&
  {Carlberg}}{1991}]{Dubinski1991}
{Dubinski} J.,  {Carlberg} R.~G.,  1991, \apj, 378, 496

\bibitem[\protect\citeauthoryear{{Fasano}, {Amico}, {Bertola}, {Vio} \&
  {Zeilinger}}{{Fasano} et~al.}{1993}]{Fasano1993}
{Fasano} G.,  {Amico} P.,  {Bertola} F.,  {Vio} R.,    {Zeilinger} W.~W.,
  1993, \mnras, 262, 109

\bibitem[\protect\citeauthoryear{{Franx}, {van Gorkom} \& {de Zeeuw}}{{Franx}
  et~al.}{1994}]{Franx+1994}
{Franx} M.,  {van Gorkom} J.~H.,    {de Zeeuw} T.,  1994, \apj, 436, 642

\bibitem[\protect\citeauthoryear{{Frenk}, {White}, {Davis} \&
  {Efstathiou}}{{Frenk} et~al.}{1988}]{Frenk1988}
{Frenk} C.~S.,  {White} S.~D.~M.,  {Davis} M.,    {Efstathiou} G.,  1988, \apj,
  327, 507

\bibitem[\protect\citeauthoryear{{Gadotti} \& {de Souza}}{{Gadotti} \& {de
  Souza}}{2003}]{GadottideSouza2003}
{Gadotti} D.~A.,  {de Souza} R.~E.,  2003, \apjl, 583, L75

\bibitem[\protect\citeauthoryear{{Gerhard} \& {Vietri}}{{Gerhard} \&
  {Vietri}}{1986}]{GerhardVietri1986}
{Gerhard} O.~E.,  {Vietri} M.,  1986, \mnras, 223, 377

\bibitem[\protect\citeauthoryear{{Hayashi}, {Navarro} \& {Springel}}{{Hayashi}
  et~al.}{2007}]{Hayashi2007}
{Hayashi} E.,  {Navarro} J.~F.,    {Springel} V.,  2007, \mnras, 377, 50

\bibitem[\protect\citeauthoryear{{Heller}, {Shlosman} \&
  {Athanassoula}}{{Heller} et~al.}{2007}]{HellerShlosmanAthanassoula2007a}
{Heller} C.~H.,  {Shlosman} I.,    {Athanassoula} E.,  2007, \apj, 671, 226

\bibitem[\protect\citeauthoryear{{Hernquist}}{{Hernquist}}{1993}]{Hernquist199%
3}
{Hernquist} L.,  1993, \apjs, 86, 389

\bibitem[\protect\citeauthoryear{{Holley-Bockelmann}, {Mihos}, {Sigurdsson} \&
  {Hernquist}}{{Holley-Bockelmann} et~al.}{2001}]{HolleyBockelmann2001}
{Holley-Bockelmann} K.,  {Mihos} J.~C.,  {Sigurdsson} S.,    {Hernquist} L.,
  2001, \apj, 549, 862

\bibitem[\protect\citeauthoryear{{Jing} \& {Suto}}{{Jing} \&
  {Suto}}{2002}]{Jing2002}
{Jing} Y.~P.,  {Suto} Y.,  2002, \apj, 574, 538

\bibitem[\protect\citeauthoryear{{Jog}}{{Jog}}{2000}]{Jog2000}
{Jog} C.~J.,  2000, \apj, 542, 216

\bibitem[\protect\citeauthoryear{{Kazantzidis}, {Kravtsov}, {Zentner},
  {Allgood}, {Nagai} \& {Moore}}{{Kazantzidis} et~al.}{2004}]{Kazantzidis2004}
{Kazantzidis} S.,  {Kravtsov} A.~V.,  {Zentner} A.~R.,  {Allgood} B.,  {Nagai}
  D.,    {Moore} B.,  2004, \apjl, 611, L73

\bibitem[\protect\citeauthoryear{{Lambas}, {Maddox} \& {Loveday}}{{Lambas}
  et~al.}{1992}]{Lambas1992}
{Lambas} D.~G.,  {Maddox} S.~J.,    {Loveday} J.,  1992, \mnras, 258, 404

\bibitem[\protect\citeauthoryear{{Maciejewski} \& {Athanassoula}}{{Maciejewski}
  \& {Athanassoula}}{2007}]{MaciejewskiAthanassoula2007}
{Maciejewski} W.,  {Athanassoula} E.,  2007, \mnras, 380, 999

\bibitem[\protect\citeauthoryear{{Maciejewski} \& {Sparke}}{{Maciejewski} \&
  {Sparke}}{2000}]{MS2000}
{Maciejewski} W.,  {Sparke} L.~S.,  2000, \mnras, 313, 745

\bibitem[\protect\citeauthoryear{{Martinez-Valpuesta}, {Shlosman} \&
  {Heller}}{{Martinez-Valpuesta} et~al.}{2006}]{Martinez-Valpuesta+2006}
{Martinez-Valpuesta} I.,  {Shlosman} I.,    {Heller} C.,  2006, \apj, 637, 214

\bibitem[\protect\citeauthoryear{{Novak}, {Cox}, {Primack}, {Jonsson} \&
  {Dekel}}{{Novak} et~al.}{2006}]{Novak+2006}
{Novak} G.~S.,  {Cox} T.~J.,  {Primack} J.~R.,  {Jonsson} P.,    {Dekel} A.,
  2006, \apjl, 646, L9

\bibitem[\protect\citeauthoryear{{O'Neill} \& {Dubinski}}{{O'Neill} \&
  {Dubinski}}{2003}]{ONeilDubinski2003}
{O'Neill} J.~K.,  {Dubinski} J.,  2003, \mnras, 346, 251

\bibitem[\protect\citeauthoryear{{Raha}, {Sellwood}, {James} \& {Kahn}}{{Raha}
  et~al.}{1991}]{Raha1991}
{Raha} N.,  {Sellwood} J.~A.,  {James} R.~A.,    {Kahn} F.~D.,  1991, \nat,
  352, 411

\bibitem[\protect\citeauthoryear{{Rix} \& {Zaritsky}}{{Rix} \&
  {Zaritsky}}{1995}]{RixZaritsky1995}
{Rix} H.-W.,  {Zaritsky} D.,  1995, \apj, 447, 82

\bibitem[\protect\citeauthoryear{{Rodionov}, {Athanassoula} \&
  {Sotnikova}}{{Rodionov} et~al.}{2009}]{Rodionov2009}
{Rodionov} S.~A.,  {Athanassoula} E.,    {Sotnikova} N.~Y.,  2009, \mnras, 392,
  904

\bibitem[\protect\citeauthoryear{{Ryden}}{{Ryden}}{2004}]{Ryden2004}
{Ryden} B.~S.,  2004, \apj, 601, 214

\bibitem[\protect\citeauthoryear{{Ryden}}{{Ryden}}{2006}]{Ryden2006}
{Ryden} B.~S.,  2006, \apj, 641, 773

\bibitem[\protect\citeauthoryear{{Sellwood}}{{Sellwood}}{1980}]{Sellwood1980}
{Sellwood} J.~A.,  1980, \aap, 89, 296

\bibitem[\protect\citeauthoryear{{Tissera}, {White}, {Pedrosa} \&
  {Scannapieco}}{{Tissera} et~al.}{2009}]{Tissera+2009}
{Tissera} P.~B.,  {White} S.~D.~M.,  {Pedrosa} S.,    {Scannapieco} C.,  2009,
  arXiv:0911.2316

\bibitem[\protect\citeauthoryear{{Trachternach}, {de Blok}, {Walter}, {Brinks}
  \& {Kennicutt}}{{Trachternach} et~al.}{2008}]{Trachternach2008}
{Trachternach} C.,  {de Blok} W.~J.~G.,  {Walter} F.,  {Brinks} E.,
  {Kennicutt} R.~C.,  2008, \aj, 136, 2720

\bibitem[\protect\citeauthoryear{{Valluri}, {Debattista}, {Quinn} \&
  {Moore}}{{Valluri} et~al.}{2009}]{Valluri2009arXiv}
{Valluri} M.,  {Debattista} V.~P.,  {Quinn} T.,    {Moore} B.,  2009,
  arXiv:0906.4784

\bibitem[\protect\citeauthoryear{{Villa-Vargas}, {Shlosman} \&
  {Heller}}{{Villa-Vargas} et~al.}{2009}]{VillaVargas+2009}
{Villa-Vargas} J.,  {Shlosman} I.,    {Heller} C.,  2009, \apj, 707, 218

\bibitem[\protect\citeauthoryear{{Warren}, {Quinn}, {Salmon} \&
  {Zurek}}{{Warren} et~al.}{1992}]{Warren1992}
{Warren} M.~S.,  {Quinn} P.~J.,  {Salmon} J.~K.,    {Zurek} W.~H.,  1992, \apj,
  399, 405

\bibitem[\protect\citeauthoryear{{Widrow}}{{Widrow}}{2008}]{Widrow2008}
{Widrow} L.~M.,  2008, \apj, 679, 1232

\end{thebibliography}

\bsp

\label{lastpage}

\end{document}